\documentclass[12pt]{article}

\usepackage[dvips]{graphicx}
\usepackage{amsmath}
\usepackage{amsfonts}
\usepackage{afterpage} 

\usepackage[a4paper]{typearea} 
  \areaset[0cm]{17.0cm}{24cm}       

\newcommand{\PSImagx}[2]{\includegraphics[width=#2]{rpsplots/#1}} 
 
\newcommand{\PSImagxrotated}[3]{%
\includegraphics[width=#2,angle=#3]{rpsplots/#1}}

\newcommand{\BILD}[4]{\begin{figure}[#1]%

     #2

     \centerline{\parbox{15cm}{\caption[.]{#3} \label{#4}}}
     \end{figure} }

\newcommand{\Int}{\int\limits}
\newcommand{\IInt}{\iint\limits}

\newcommand{\ud}{\text{d}}
\newcommand{\ui}{\text{i}}
\newcommand{\ue}{\text{e}}

\newcommand{\Vol}{\operatorname{vol}}

\newcommand{\R}{\mathbb{R}}

\newcommand{\N}{\mathbb{N}}

\begin{document}


\vspace{0.5cm}

\vspace*{1cm}

\normalsize

\vspace{0.5cm}

\renewcommand{\thefootnote}{\fnsymbol{footnote}}

\begin{center}
  {\huge\bf Chaotic eigenfunctions \\ in momentum space}
\end{center}

\setcounter{footnote}{1}

\begin{center}
   \vspace{3ex}
 
         {\large A.\ B\"acker%
\footnote[1]{E-mail address: {\tt baec@physik.uni-ulm.de}}
          and R.\ Schubert%
\footnote[7]{E-mail address: {\tt schub@physik.uni-ulm.de}}
          }

   \vspace{4ex}

   Abteilung Theoretische Physik\\
   Universit\"at Ulm\\
   Albert-Einstein-Allee 11\\
   D--89069 Ulm\\
   Federal Republic of Germany\\

\end{center}

\renewcommand{\thefootnote}{\arabic{footnote}}
\setcounter{footnote}{0}


\newcommand{\p}{\partial}
\newcommand{\Ci}{\operatorname{Ci}}
\newcommand{\Si}{\operatorname{Si}}
\newcommand{\Ha}{\operatorname{\mathbf{H}}}
\newcommand{\fp}{\operatorname{FP}}

\newcommand{\EE}{\Sigma_E}

\newcommand{\omm}{s}   
\newcommand{\lx}{l_x}
\newcommand{\ly}{l_y}

\newcommand{\BFx}{\boldsymbol q}
\newcommand{\BFq}{\BFx}
\newcommand{\BFp}{\boldsymbol p}
\newcommand{\BFn}{\boldsymbol n}
\newcommand{\noBFx}{{q}}
\newcommand{\noBFq}{{\noBFx}}
\newcommand{\noBFp}{{p}}

\vspace*{3cm}
\leftline{\bf Abstract:}

\vspace*{0.1cm}
{\small 
We study eigenstates of chaotic billiards 
in the momentum representation and propose the radially integrated momentum
distribution as useful measure to detect localization effects.
For the momentum distribution, the radially integrated
momentum distribution, and the angular integrated momentum
distribution explicit formulae in terms of the normal derivative
along the billiard boundary are derived.
We present a detailed numerical study 
for the stadium and the cardioid billiard, which shows
in several cases that the radially integrated momentum distribution 
is a good indicator of localized eigenstates,
such as scars, or bouncing ball modes.
We also find examples, where the localization
is more strongly pronounced in position space than in momentum space,
which we discuss in detail.
Finally applications and generalizations are discussed.

}

\newpage

\section{Introduction}

In quantum mechanics the state of a system is given by a normalized vector in 
a Hilbert space, which in turn can be represented in  different ways. 
Usually one chooses the representation which is most convenient 
for the problem at hand. In quantum chaos the 
main concern is to understand the 
semiclassical limit, and the fingerprints which properties of the 
classical limit, like chaoticity or integrability, leave on 
the quantum system. 

A particulary convenient class of representations for the study of the 
quantum to classical correspondence is given by the Wigner function  and its 
relatives, e.g.\ the Husimi density. 
The Wigner function of a quantum state 
is the representation which comes closest to a probability density 
on the phase space 
of the corresponding classical system \cite{Ber77a,HilConScuWig84}, and 
therefore should be a sensitive detector for classical 
fingerprints. One drawback of the Wigner function is that 
it is difficult 
to visualize, because it is a function on the $2n$--dimensional 
phase space. Therefore one is often forced to study the projections 
on position or momentum space when one wants a visual representation 
of the state. Whereas the position representation 
is commonly used in quantum chaos, the momentum representation is
not. Our aim here is to promote the momentum 
representation, by showing that it has some potential advantages. 

What we are especially looking for are fingerprints of the 
classical phase space 
structure, e.g.\ periodic orbits. So in the position representation one
is searching for an enhanced probability density around an orbit, in 
order to detect e.g.\ so-called scarred eigenstates \cite{Hel84}. But the     
detection of such  a state  from 
the probability density in position space can be rather ambiguous, there is 
no clear borderline which distinguishes scarred eigenstates 
from those which are not.
In this case the probability density in momentum space can be helpful. 
E.g.\ assume we are studying an Euclidean billiard system, 
where the periodic orbits consist of segments of straight lines, 
therefore the momentum distribution of a state which is scarred 
by a periodic orbit should be concentrated on the directions of the 
line segments of that orbit. This criterion can be 
further simplified if one realizes that the momentum distribution 
of the n'th state is concentrated on the energy shell $\BFp^2=E_n$, so that 
the main information is contained in the dependence on the direction of 
$\BFp$. 
By integrating over the radial direction one gets a one--dimensional 
distribution, the angular distribution of the momentum, which 
contains all the essential information about the momentum distribution. 
This is the main advantage of 
the momentum representation, instead of working with a two--dimensional 
distribution as in position space, one can reduce 
the study to the one--dimensional angular distribution of the momentum.

A related quantity was introduced in \cite{KlaSmi96}, where 
the scattering approach is used to study expectation values and scars 
on the Poinc{\'a}re section. The expectation values of a
proposed ``scar operator''
correspond to a smoothed angular distribution of the momentum. 
Other approaches to study eigenfunctions consist in the use of
the Bargmann representation and the stellar representation \cite{LebVor90}.
For billiards these can be used to obtain representations
on the Poincar\'e section \cite{TuaVor95}, see also \cite{CrePerCha93}.

The plan of the paper is the following. In the first section we 
use the boundary integral method to obtain expressions for 
the momentum representation of a given eigenfunction in terms of 
the normal derivative of this 
eigenfunction on the boundary. Furthermore we derive
formulas for the angular and the radial 
distribution of the momentum density, respectively,  expressed in 
terms of the normal derivative. 
In the second section 
we present a gallery of eigenstates displayed in position space
and in momentum space, and compare the different representations. 
Finally applications and generalizations are discussed.

\section{Momentum distributions}

The boundary integral method is a common method for computing 
eigenvalues and eigenfunctions of two--dimensional billiards,
see e.g.\ \cite{Rid79,BerWil84}.
The main point is that one can reduce the two--dimensional 
eigenvalue problem in such a billiard with the help of  
Green's formula and a Green function into an integral equation for the 
normal derivative of the eigenfunction on the boundary. Therefore 
one needs to solve only a one--dimensional problem, 
which is much more efficient numerically.

\subsection{Reduction to the boundary}

Let $\Omega\subset \R^2$ be a domain in the Euclidean plane with 
piecewise smooth boundary $\p\Omega$. The Dirichlet quantum billiard
in $\Omega$ is defined by 
\begin{align}
  -\Delta \psi_n (\BFx)&= p_n^2\psi_n (\BFx)& 
                                      \text{for}&\quad 
                            \BFx\in \Omega \backslash \p\Omega \\
   \psi_n(\BFx)        &= 0&      \text{for}&\quad \BFx\in \p\Omega \,\, .
\end{align}
And we will use the trivial extension of $\psi_n(\BFx)$ to 
$\R^2$ by requiring $\psi_n(\BFx)=0$ for 
$\BFx\in \R^2\setminus \Omega$ to $\R^2$. The transformation 
of this problem to a problem on the boundary rests on the 
formula of Green, 
\begin{equation}\label{green}
  \Int_{\Omega}\bigl[f(\BFx)\Delta g(\BFx)- g(\BFx)\Delta f(\BFx)
  \bigr]\;\! \ud^2\noBFx =\! 
  \Int_{\p\Omega}\bigl[f(\BFx(\omm) )\BFn(\omm )
  \nabla g(\BFx(\omm) )-g(\BFx(\omm))\BFn(\omm )
  \nabla f(\BFx(\omm) )\bigr]\;\! \ud\omm \; ,
\end{equation}
where $f$ and $g$ are arbitrary smooth functions, and 
$\BFn(\omm )$ denotes the outer normal on $\p \Omega$, which is 
defined almost everywhere if the boundary is piecewise smooth. 
If one inserts for $f$  a Green function, e.g.\
$f(\BFx)=\frac{\ui}{4}H_0^{(1)}(p_n|\BFx-\BFx' |)$,
where  $H_0^{(1)}=J_0+\ui Y_0$ is the Hankel function 
of the first kind,  which satisfies 
\begin{equation}
  \left(\Delta_{\BFx} +p_n^2\right)\frac{\ui}{4}H_0^{(1)}(p_n|\BFx-\BFx' |)
   =\delta (\BFx-\BFx' )\;\; ,
\end{equation}
and for $g$ the eigenfunction $\psi_n$, one gets for 
$\BFx\in \R^2\setminus\p\Omega$ 
\begin{equation}\label{bint0}
  \psi_n(\BFx) =\Int_{\p\Omega}\frac{\ui}{4}H_0^{(1)}(p_n|\BFx(\omm )-\BFx |) 
          \,u_n (\omm )\; \ud\omm \;\; ,
\end{equation}
where $u_n (\omm ):=\BFn (\omm) \nabla\psi_n (\BFx(\omm ) ) $ 
denotes the normal derivative of $\psi_n$ on the boundary expressed
in terms of the arclength coordinate $s$. 
Eq.~\eqref{bint0} is the crucial relation by which one recovers the 
eigenfunction from its normal derivative on the boundary. Taking the 
normal derivative of \eqref{bint0} leads to an integral equation for $u_n$ 
on the boundary, which is then solved numerically. 
The choice 
of $H_0^{(1)}$ for the Green function is necessary in order that  
this integral equation for $u_n$  does not lead to any spurious 
solutions, see e.g.\ \cite{Rid79}. For the integral 
(\ref{bint0}) the contribution of $J_0$ is irrelevant, because
with  $(\Delta +p_n^2)J_0(p_n|\BFx|)=0$ the Green formula gives 
for $\BFx\in \Omega\setminus\p\Omega$ 
\begin{equation}
  \Int_{\p\Omega} J_0(p_n|\BFx(\omm )-\BFx |) \, u_n (\omm )\; \ud\omm =0\;\; ,
\end{equation}
and so one has the simpler formula for the eigenfunction 
for $\BFx\in \Omega\setminus\p\Omega$ 
\begin{equation}\label{bint}
  \psi_n(\BFx) =-\frac{1}{4}\Int_{\p\Omega}
            Y_0(p_n|\BFx(\omm )-\BFx |)\,u_n (\omm )\; \ud\omm \;\; .
\end{equation}
Furthermore, one can express the $L_2$-norm of $\psi_n$ by $u_n$ as 
\cite{BerWil84,Boa94}
\begin{equation} \label{eq:boundary-normalization}
  \frac{1}{2p_n^2}\Int_{\p\Omega }\BFn(\omm )\BFx(\omm ) \, 
       | u_n (\omm )|^2\, \ud\omm = ||\psi_n ||^2 \;\; .
\end{equation}
This expression is very convenient as it allows
to obtain a normalized normal derivative $u_n$ 
such that all other derived quantities of interest, such as expectation
values, are correctly normalized. 
This relation is much faster to compute numerically than
a normalization obtained from integrating $|\psi(\BFx)|^2$ over
the billiard domain $\Omega$.

Even in cases where the eigenfunctions are given by an expansion
into some basis it can be useful to perform the
computations starting from the normal derivative, as one
can exploit the relation \eqref{eq:boundary-normalization}
to obtain normalized eigenfunctions.

\subsection{Momentum representation of eigenfunctions}

The momentum representation of an eigenfunction $\psi_n$
is given by its Fourier transform 
\begin{equation}
  \widehat{\psi}_n(\BFp):=\frac{1}{2\pi}\Int_{\R^2}\ue^{-\ui\BFp\BFx}
  \psi_n (\BFx)\, \ud^2\noBFx 
  =\frac{1}{2\pi}\Int_{\Omega}\ue^{-\ui \BFp\BFx}\psi_n (\BFx)\, 
    \ud^2\noBFx \;\; .
\end{equation}
Using Green's formula (\ref{green}) with $f(\BFx)=\psi_n(\BFx)$ and 
$g(\BFx)=\frac{1}{2\pi}\ue^{-\ui \BFx\BFp}$ one gets 
\begin{equation} 	 
\begin{split}
  \frac{1}{2\pi}
  \Int_{\Omega}&\left[\psi_n(\BFx)(-\ui \BFp)^2 \,\ue^{-\ui \BFx\BFp}-
  \ue^{-\ui \BFx\BFp}(-p_n^2)\psi_n(\BFx)\right]\; \ud^2\noBFx \\
 & =\frac{1}{2\pi}\Int_{\p\Omega}
  \left[\psi_n(\BFx(\omm))(-\ui \BFn(\omm)\BFp)
      \ue^{-\ui \BFx(\omm)\BFp}-u_n(\omm ) 
      \ue^{-\ui \BFx(\omm)\BFp}\right]\; \ud\omm \, . 
\end{split} 
\end{equation}
As $\psi_n(\BFx)=0$ for $\BFx\in \p\Omega$ we get a representation 
of the Fourier transform as an integral over the normal derivative at the 
boundary
\begin{equation}\label{ftrans}
  \widehat{\psi}_n(\BFp)=\frac{1}{\BFp^2-p_n^2}\frac{1}{2\pi}\Int_ {\p\Omega}
  u_n(\omm )\ue^{-\ui \BFx(\omm )\BFp}\, \ud\omm \;\; .
\end{equation}
This expression has an apparent singularity at 
$\BFp^2=p_n^2$, but since  the Fourier transform 
and its derivatives are bounded
this singularity has to be cancelled by a zero of the integral. Solving 
\eqref{ftrans} for the integral gives  
\begin{equation}
\frac{1}{2\pi}\Int_ {\p\Omega}
  u_n(\omm )\ue^{-\ui \BFx(\omm)\BFp}\, \ud\omm =\left(\BFp^2-p_n^2\right)
      \widehat{\psi}_n(\BFp) \;\; ,
\end{equation}
and taking the derivative of this equation with respect to $r:=|\BFp|$ at 
the point $r=p_n$ one gets 
\begin{equation}
  \widehat{\psi}_n(\BFp)=-\frac{\ui}{4\pi p_n^2}
       \Int_{\p\Omega}\ue^{-\ui \BFp\BFx(\omm )}  
  \BFp\BFx(\omm )u_n(\omm )\,\ud\omm \, ,
\end{equation}
for $|\BFp|=p_n$. By repeating this procedure one can get expressions for the 
derivatives of $\widehat{\psi}_n(\BFp)$ at $|\BFp|=p_n$ as well. 
This can be 
used to interpolate near $|\BFp|=p_n$
for a numerical plot of $\widehat{\psi}_n(\BFp)$.

\subsection{Radially integrated momentum distribution}

In a billiard the  classical flows at different energies $\BFp^2$ 
are isomorphic, because the system scales with energy. 
At the quantum mechanical side the momentum distribution of 
the $n$-th state is concentrated around the energy shell 
$\BFp^2=p_n^2$.
Because of the scaling property of 
the classical system the interesting phenomena, i.e.\ those 
which are due to the special system at hand, should occur only 
in the angular distribution of the momentum-probability.

Therefore we want to study the angular distribution of the 
momentum defined by 
\begin{equation}\label{Idef}
  I_n(\varphi ):=\Int_0^{\infty}\left|\widehat{\psi}_n(r,\varphi )
                               \right|^2 r\; \ud r \;\; , \end{equation}   
where we have introduced polar coordinates 
$\BFp=(r\cos\varphi ,r\sin\varphi )$. The advantage 
of this quantity, in contrast to the full momentum probability density or 
the probability distribution in position space, is that it 
depends only on one variable $\varphi\in [0,2\pi ]$.  

By inserting the expression (\ref{ftrans}) for the Fourier transform 
$\widehat{\psi}_n$ into (\ref{Idef}) one gets 
\begin{equation}
  I_n(\varphi )=\frac{1}{(2\pi)^2}\Int_0^{\infty}\frac{r}{(r^2-p_n^2)^2} 
  \;\;
  \IInt_{\p\Omega\times\p\Omega }\ue^{-\ui r\alpha }\,u_n(\omm )
         \overline{u}_n(\omm ' )
  \; \ud\omm \,\ud\omm ' \; \ud r \;\; ,
\end{equation}
where we have used the abbreviation 
\begin{equation}
\label{alphadef}
 \alpha = \alpha(\varphi ,\omm,\omm')
       :=\hat{\BFp}(\varphi )(\BFx(\omm )-\BFx(\omm '))\;\; ,
\end{equation}
 and $\hat{\BFp}(\varphi )=(\cos\varphi ,\sin\varphi )$ is the 
unit vector in $\BFp$ direction. Using
$ \frac{r}{(r^2-p_n^2)^2} =-\frac{1}{2}\frac{\ud}{\ud r}\frac{1}{r^2-p_n^2}$
and partial integration one obtains
\begin{equation}\label{partintI}
  4\pi^2 I_n(\varphi )=-2\pi^2 p_n^2\left|\widehat{\psi}_n(0)\right|^2 \,\,
  -\frac{1}{2}\Int_0^{\infty}\frac{1}{r^2-p_n^2}
  \;\;\IInt_{\p\Omega\times\p\Omega }\ui\alpha \,
  \ue^{-\ui r\alpha }u_n(\omm )\overline{u}_n(\omm ' )
  \; \ud\omm \,\ud\omm ' \; \ud r \;\; ,
\end{equation}
where it was used that $\int_{\p\Omega}u_n(\omm )\, \ud\omm =-2\pi p_n^2 
\widehat{\psi}_n(0)$ which follows from (\ref{ftrans}).

Here we have to discuss a problem which always occurs  when  
inserting the expression (\ref{ftrans}) for $\widehat{\psi}_n$ in an integral 
such as in eq.~(\ref{Idef}). Since (\ref{ftrans}) is the product of a 
factor which becomes singular at $|p|=p_n$ and a factor which is zero 
there, we have an apparent problem if we want to interchange the 
order of integration in  (\ref{partintI}). Therefore we will choose a 
regularization for the factor $\frac{1}{\BFp^2-p_n^2}$. This can be done 
by adding to  $p_n$ a small positive or negative imaginary part, 
 $p_n\pm \ui\epsilon$, performing all computations 
and taking afterwards the limit $\epsilon \to 0$. The result is 
independent of the regularization. In the following we will choose 
a symmetric regularization, i.e.,
replace $(\BFp^2-p_n^2)^{-1}$ by 
\begin{equation}
  \frac{1}{2}\left(\frac{1}{\BFp^2-(p_n+\ui\epsilon)^2}
  +\frac{1}{\BFp^2-(p_n-\ui\epsilon)^2}\right)\;\; ,
\end{equation}
which corresponds in the limit 
$\epsilon \to 0$ to taking the principal values of all integrals over 
$r=|\BFp|$. To avoid  complicated formulas we will proceed 
formally without regularization, and interchange integrals whenever 
needed, but interpret all integrals over $r$ as 
principal values. The reader can always check that the 
described regularization leads to the same results. 

Returning to the computation of $I_n(\varphi )$ we interchange 
the order of integration in (\ref{partintI}) and get 
\begin{equation}
  I_n(\varphi )=\IInt_{\p\Omega\times\p\Omega } \,K_n(\varphi ,\omm ,\omm ')
  \,u_n(\omm )\overline{u}_n(\omm ' )
  \; \ud\omm \ud\omm' \;\;, 
\end{equation}
where the kernel is given by 
\begin{equation}
\begin{split}
  K_n(\varphi ,\omm ,\omm ')
  &=-\frac{\ui\alpha }{8\pi^2}
   \Int_0^{\infty}\frac{\ue^{-\ui r\alpha }}{r^2-p_n^2}\;  \ud r
     -\frac{1}{2p_n^2}\\
  &=-\frac{\alpha }{8\pi^2}
  \Int_0^{\infty}\frac{\sin(r\alpha )}{r^2-p_n^2}\; \ud r-\frac{1}{2p_n^2}
  -\ui\frac{\alpha }{8\pi^2}
        \Int_0^{\infty}\frac{\cos(r\alpha )}{r^2-p_n^2}\; \ud r \;\;, 
\end{split}
\end{equation}
and the integrals are interpreted as  principal values. Note that
the real part is an even function of $\alpha$ and the imaginary part is 
an odd function of $\alpha$, thus $K_n$ has the property 
\begin{equation}
  K_n(\varphi ,\omm ,\omm ')=\overline{K}_n(\varphi ,\omm ' ,\omm ) \;\;, 
\end{equation}
which is equivalent to $\overline{I}_n(\varphi ) =I_n(\varphi )$. 
For Euclidean billiards the eigenfunctions can be choosen to be real. 
In this case  $\overline{u}_n(\omm )u_n(\omm ')$ is 
even under interchange of $\omm$ and $\omm '$, and therefore the integral 
over the imaginary part of $K_n$, which is odd, vanishes.

From \cite{GroHof73} one gets  for the integrals (interpreted as 
principal values)
\begin{align}
  -\frac{\alpha }{8\pi^2}
    \Int_0^{\infty}\frac{\sin(r\alpha )}{r^2-p_n^2}\, \ud r
  &=\frac{1}{8\pi^2}
    \left(\sin (|\alpha |p_n)\Ci (|\alpha |p_n)-
    \cos (|\alpha |p_n)\Si (|\alpha |p_n)\right)\\
    -\frac{\alpha }{8\pi^2}\Int_0^{\infty}
               \frac{\cos(r\alpha )}{r^2-p_n^2}\, \ud r
  &=\frac{\alpha }{16\pi p_n} \sin (|\alpha |p_n)\, \, ,
\end{align}
where $\Si$ and $\Ci$ denote the sine and cosine integrals. 
Note, that for billiards with Dirichlet boundary conditions 
$p_n>0$. With the abbreviation 
\begin{equation}
  f(x):=\sin(x)\Ci (x)-\cos (x) \Si (x) \;\;, 
\end{equation}
we can write our final expression for $I_n(\varphi )$ in terms 
of the normal derivative $u_n$ as 
\begin{equation}\label{Ierg}
\begin{split}
  I_n(\varphi )=\IInt_{\p\Omega\times\p\Omega }
  &\left[\frac{1}{8\pi^2}f(|\alpha |p_n)
    -\frac{1}{2p_n^2}\right]
    \,u_n(\omm )\,\overline{u}_n(\omm ' )\; \ud\omm\, \ud\omm'\\
  & +
    \ui\IInt_{\p\Omega\times\p\Omega }\frac{\alpha }{16\pi p_n} 
    \sin (|\alpha |p_n)
    \,u_n(\omm )\,\overline{u}_n(\omm ' )\; \ud\omm \,\ud\omm'\;\; .
\end{split}
\end{equation}
Recall that $\alpha=\alpha(\varphi,\omm,\omm')=
\hat{\BFp}(\varphi )(\BFx(\omm )-\BFx(\omm '))$. 
If the normal derivative $u_n$ is real
then the second term in (\ref{Ierg}) vanishes, and if furthermore the 
eigenfunction $\psi_n$ is odd with respect to some discrete 
symmetry of the billiard, see the appendix, 
then $\int_{\p\Omega }u_n(\omm )\, \ud\omm =0$. 
So under these special conditions the expression (\ref{Ierg}) 
simplifies to 
\begin{equation} 
  I_n(\varphi )=\frac{1}{8\pi^2}\IInt_{\p\Omega\times\p\Omega }
  f(|\alpha |p_n)u_n(\omm )u_n(\omm ')\;\ud\omm \,\ud\omm'\;\; .
\end{equation}

\subsection{Angular integrated momentum density}

Similarly to the radially  integrated momentum density 
one can study the radial distribution 
of the momentum density by integrating over the angle $\varphi$. 
The radial momentum distribution is defined as 
\begin{equation}
R_n(r):=r\Int_0^{2\pi}\left|\widehat{\psi}_n(r,\varphi )\right|^2\,\, 
\ud \varphi
\end{equation}
and if we insert  the expression (\ref{ftrans}) for 
$\widehat{\psi}_n$  one gets 
\begin{align}
  R_n(r):&=\frac{r}{(r^2-p_n^2)^2}\frac{1}{(2\pi )^2}
  \IInt_{\p\Omega\times\p\Omega }u_n(\omm )\,\overline{u}_n(\omm ' )
  \Int_0^{2\pi}\ue^{-\ui|\BFx(\omm )-\BFx(\omm ' )|
            r\cos(\varphi)}\,\,\ud \varphi 
  \; \ud\omm \,\ud\omm '\notag\\
  &=\frac{r}{(r^2-p_n^2)^2}\frac{1}{2\pi } 
  \IInt_{\p\Omega\times\p\Omega }u_n(\omm )\,\overline{u}_n(\omm ' )
  J_0(|\BFx(\omm )-\BFx(\omm ' )|r)  \; \ud\omm \,\ud\omm '\,\, .\label{Pn}
\end{align}
Again one has an apparent singularity at $r=p_n$ due to the 
pre--factor $1/(r^2-p_n^2)^2$.
In the same 
way as in the previous section one obtains by differentiating 
$(r^2-p_n^2)^2R_n(r)/r$ sufficiently often 
\begin{align}
  R_n(p_n)&=\frac{1}{8p_n}\left(\frac{\ud}{\ud r}\right)^2
  {\left((r^2-p_n^2)^2\frac{R_n(r)}{r}\right)}_{r=p_n}\\
  R_n'(p_n)&=\frac{1}{24p_n}\left(\frac{\ud}{\ud r}\right)^3
  {\left((r^2-p_n^2)^2\frac{R_n(r)}{r}\right)}_{r=p_n}\,\, ,
\end{align}
and higher derivatives of $R_n$ at $r=p_n$ can be obtained in the same way. 
Using eq.~(\ref{Pn}) one gets 
\begin{align}
  R_n(p_n)&=
  \frac{1}{16\pi p_n } 
  \IInt_{\p\Omega\times\p\Omega }u_n(\omm )\,\overline{u}_n(\omm ' )
  |\BFx(\omm )-\BFx(\omm ' )|^2\notag \\
  &\qquad\qquad\qquad\qquad \frac{1}{2}\left[J_2(|\BFx(\omm )-\BFx(\omm ' )|r)
      -J_0(|\BFx(\omm )-\BFx(\omm ' )|r)\right] \; \ud\omm \,\ud\omm ' \\ 
  R_n'(p_n)&=\frac{1}{48\pi p_n } 
  \IInt_{\p\Omega\times\p\Omega }u_n(\omm )\,\overline{u}_n(\omm ' )
  |\BFx(\omm )-\BFx(\omm ' )|^3\notag \\
  &\qquad\qquad\qquad\qquad \frac{1}{4}
         \left[ 3 J_1(|\BFx(\omm )-\BFx(\omm ' )|r)
                - J_3(|\BFx(\omm )-\BFx(\omm ' )|r)
          \right]  \; \ud\omm \,\ud\omm ' \,\, ,
\end{align}
and these expressions can be used to interpolate in the region around $r=p_n$.

\section{Gallery of eigenfunctions in momentum space}

In the following we present a number of examples of eigenfunctions
in momentum representation for two chaotic billiards
and compare them with the position representation.
A series of eigenfunctions 
of the cosine billiard  in momentum representation 
can be found in \cite{Sti96:PhD}.
The first system we study 
is the stadium billiard, which is proven to be strongly
chaotic, i.e.\ it is ergodic, mixing and a $K$-system \cite{Bun74,Bun79}.
The height of the desymmetrized billiard 
is chosen to be 1, and $a$ denotes the length
of the upper horizontal line, for which we have $a=1.8$
in the following.
To compare the structures in the eigenfunctions 
with the classical orbits, we use the symbolic dynamics
proposed in \cite{BihKva92}, which is proven in \cite{BaeChe98}.

The second system is the cardioid billiard, which
is the limiting case of a family of billiards introduced in \cite{Rob83}.
The cardioid billiard is proven to be ergodic, mixing, a $K$-system
and a Bernoulli system \cite{Woj86,Sza92,Mar93}.
The eigenvalues of the cardioid billiard have been provided
by Prosen and Robnik \cite{PrivComProRob} and were calculated
by means of the conformal mapping technique, see e.g.\ 
\cite{Rob84,ProRob93a}.
To describe the periodic orbits in the cardioid
we use the symbolic dynamics proposed in \cite{BruWhe96,BaeDul97},
see \cite{BaeChe98,Dul98b} for proofs.

As these two billiards are ergodic, the quantum ergodicity theorem 
\cite{Shn74,Shn93,Zel87,CdV85,HelMarRob87} applies, 
see e.g.\ \cite{Sar95,KnaSin97,BaeSchSti98} for introductions.
It implies that for ``almost all'' eigenfunctions their Wigner functions 
$W_n(\BFp,\BFq)$ 
become equidistributed on the energy shell in the semiclassical limit, i.e.\
\begin{equation} \label{eq:quantum-ergodicity-Wigner}
  W_{n_j}(\BFp,\BFq) := \Int_{\R^2} \ue^{ \ui \BFq' \BFp}\, 
  \, \overline{\psi}_{n_j} \left( \BFq-\frac{\BFq'}{2} \right) 
  \psi_{n_j}\left(\BFq+\frac{\BFq'}{2}\right)\,
   \ud^2 \noBFq' 
  \sim 
  \frac{\delta (\BFp^2-E_{n_j})}{\Vol (\Sigma_{E_{n_j}}) } \;\;,
\end{equation}
where $\Vol (\EE ) = \iint_{\R^2\times\Omega} 
        \delta(\BFp^2-E) \;  \ud^2 \noBFp\,\ud^2 \noBFq $ is the volume 
of the energy shell.
``Almost all'' precisely means that eq.~\eqref{eq:quantum-ergodicity-Wigner}
holds for a subsequence $\{n_j\} \subset \N$  of
density one, that is one has
\begin{equation}
  \lim_{E\to\infty} \frac{\#\{ n_j \;|\; E_{n_j} < E \}} 
{\#\{ n \;|\; E_n < E \}} = 1 \;\;.
\end{equation}

For the eigenstates in position and momentum reperesentation 
the quantum ergodicity theorem implies that
\begin{equation}
  |\psi_{n_j}(\BFq)|^2 \to \frac{1}{\Vol(\Omega)}   
   \qquad \text{ as } n_j \to \infty
\end{equation}
and
\begin{equation} 
  |\widehat{\psi}_{n_j}(\BFp)|^2 \to  \frac{1}{\pi} \delta(\BFp^2-E_{n_j})
    \qquad \text{ as } n_j \to \infty
\end{equation} 
in the weak sense.
For the radially  integrated momentum distribution it follows 
that
\begin{equation}\label{eq:I_mean}
  I_{n_j}(\varphi) \to \frac{1}{2\pi} \qquad \text{ as } n_j \to \infty
\end{equation}
in the weak sense. 
For the angular integrated momentum density one has in general, 
independent of the ergodic properties of the classical system, 
\begin{equation}
  R_{n_j}(r) \to  2 r \delta(r^2-E_{n_j})  
   \qquad \text{ as } n_j \to \infty \;\;.
\end{equation}

Of special interest are of course the subsequences of
exceptional eigenfunctions (if they exist) 
which are according to the quantum ergodicity theorem of density zero
for ergodic systems.
A drastic example of such eigenfunctions
are the so--called bouncing ball modes in the stadium (which also
occur in other billiards with two parallel walls).
These eigenfunctions localize on the orbits 
which bounce up and down between the two parallel walls
with perpendicular reflections at the boundary.
For the stadium billiard the counting function for these modes 
increases asymptotically as $E^{3/4}$ \cite{Tan97,BaeSchSti97a}.

Another important class of exceptional eigenfunctions are so--called
``scarred'' eigenfunctions \cite{Hel84}, 
showing localization along unstable periodic orbits,
which have been first observed in the stadium billiard.
Also in the cardioid billiard scarred eigenfunctions have been observed 
\cite{BaeSchSti98,Bae98:PhD}.
As has been mentioned in the introduction, for an eigenfunction 
which is scarred by a periodic orbit, $I_n(\varphi)$ should have 
prominent peaks at the angles corresponding to the directions of the orbit, 
whose intensity is expected to be proportional to the length of the 
corresponding orbit segment.

Let us start the discussion by showing plots
of some low lying eigenfunctions of the  stadium billiard
with  odd-odd symmetry. 
Figs.\ \ref{fig:stadium-ThreeD-A}--\ref{fig:stadium-ThreeD-B} 
show for the stadium billiard three--dimensional plots of 
$|\psi_n(\BFx)|^2$, 
$|\widehat{\psi}_n(\BFp)|^2$
and the corresponding grey--scale plots. Furthermore 
$I_n(\varphi)$ and $R_n(r)$ are shown.
In the case of $I_n(\varphi)$ we only plot the interval $\varphi\in[0,\pi/2]$,
as the other directions are obtained from symmetry, e.g.\
$I_n(\varphi)= I_n(\pi-\varphi)$ for $\varphi\in[\pi/2,\pi]$.
The dashed line in the plot of $I_n(\varphi)$ corresponds to the mean value 
$1/(2\pi)$, see eq.~\eqref{eq:I_mean}.
The state $n=24$ displayed in fig.~\ref{fig:stadium-ThreeD-A} 
does not show any prominent localization in position space,
whereas the state $n=26$, fig.~\ref{fig:stadium-ThreeD-B},
is an example of a low lying bouncing ball mode.
For both eigenstates the momentum distribution $|\widehat{\psi}_n(\BFp)|^2$
is mainly concentrated around the energy shell $E_n=\BFp^2$, which is 
indicated by the inner full circle in 
figs.\ \ref{fig:stadium-ThreeD-A}--\ref{fig:stadium-ThreeD-B}. 
This is also nicely seen in the plots 
of $R_{24}(r)$ and $R_{26}(r)$ where the radius of the energy 
shell is marked by a rhombus. 
Whereas $|\widehat{\psi}_{24}(\BFp)|^2$ shows peaks at several
places,  $|\widehat{\psi}_{26}(\BFp)|^2$ only has prominent
peaks around the  $p_y$ direction. 
The pictures shown in fig.~\ref{fig:stadium-ThreeD-B} 
are precisely those expected for a bouncing ball mode \cite{BaeSchSti97a}:
in position space we have localization on the rectangular
part of the billiard, and 
in momentum space there is localization in the direction of the bouncing ball
orbits.
That there are four major peaks of $|\widehat{\psi}_{26}(\BFp)|^2$
visible in fig.~\ref{fig:stadium-ThreeD-B} is due to the symmetry of 
the billiard, which implies 
that $|\widehat{\psi}_{n}(\BFp)|^2$ is reflection symmetric
with respect to the axes $p_x=0$ and $p_y=0$, on which 
$|\widehat{\psi}_{n}(\BFp)|^2$ vanishes
for the stadium billiard with everywhere Dirichlet boundary conditions.
According to \cite{BaeSchSti97a} the peaks should be in the directions  
\begin{equation} \label{eq:location-of-bb-max}
  \pm\varphi^{\text{bb}}_{k,l} 
    \quad\text{and}\quad 
  \pi\pm\varphi^{\text{bb}}_{k,l}
    \qquad\text{with}\quad 
  \varphi^{\text{bb}}_{k,l} :=\arctan(a l/k) \;\;,
\end{equation}
where $k$ denotes the number of modes in $x$--direction and 
$l$ denotes the number of modes in $y$--direction 
and $a=1.8$ is the billiard parameter. 
If the ratio $l/k$ increases, then the value of $\varphi^{\text{bb}}_{k,l}$
increases, and in the limit $l/k\to\infty$ we have 
$\varphi^{\text{bb}}_{k,l}\to\pi/2$.
For the bouncing ball mode shown in fig.~\ref{fig:stadium-ThreeD-B}
there is good agreement with the direction 
$\varphi^{\text{bb}}_{2,8} =\arctan(4 a)$,
which is marked by a full triangle in the plot of $I_{26}(\varphi)$.

Fig.~\ref{fig:stadium-two-B}a) shows an example
of an eigenfunction having a quite regular 
and uniform pattern in position space. However,
the plots of $|\widehat{\psi}_{263}(\BFp)|^2$ and $I_{263}(\varphi)$
reveal three major momentum directions.
Although we did not succeed to find an orbit
corresponding to these directions, there is some indication
that this orbit has a  quite large geometric length.

The second eigenfunction shown in fig.~\ref{fig:stadium-two-B}
is an example of a higher lying bouncing ball mode
with $14$ modes in $x$--direction and $30$ modes in $y$--direction.
For this bouncing ball mode
the distance of the maximum to these directions
is larger than for the one displayed in fig.~\ref{fig:stadium-ThreeD-B}.
This is well accounted for by the formula \eqref{eq:location-of-bb-max}
as the location of the maximum of $I_{455}(\varphi)$
agrees well with 
$\varphi^{\text{bb}}_{14,30} =\arctan(30/14\, a)$,
which is marked by a full triangle in the plot of $I_{455}(\varphi)$.

Fig.~\ref{fig:stadium-two-C} shows two examples of eigenfunctions
showing scarred structures in position space. 
$|\psi_{1771}(\BFq)|^2$ shows an enhanced probability 
along the two orbits with symbolic codes $\overline{0123}$ for the 
orbit of diamond shape,  and  
$\overline{1155}$
 for the orbit of rectangular shape (shown as dashed lines in the plot
of $|\psi_{1771}(\BFq)|^2$). 
These two structures are less visible in the momentum representations.
For the contribution of the $\overline{0123}$ orbit to $I_{1771}(\varphi)$
we find that it is spread out near to $\varphi=\pi/8$,
in agreement with the structure in position space, which is also
not aligned precisely along the $\overline{0123}$ orbit.
The higher intensity near to $\varphi=0$ 
corresponds to localization along the $\overline{1155}$ orbit.
Although $|\psi_{1771}(\BFq)|^2$ shows stronger localization near 
to the circular boundary which seems to be 
associated with the $\varphi=\pm \pi/2$ directions,
there is no prominent peak in that direction.
However, there is some enhancement of the total probability
to find the particle with momentum directions 
$3\pi/8< \varphi < \pi/2$
which is seen more clearly in the plot of 
$|\widehat{\psi}_{1771}(\BFp)|^2$ than in the plot of $I_{1771}(\varphi)$.
The rapid oscillations of $I_{1771}(\varphi)$ near
to $\varphi=\pi/2$ are not a numerical artefact; the structures
are also visible in the plot of $|\widehat{\psi}_{1771}(\BFp)|^2$, 
we will return to this point at the end of this section. 
The eigenfunction $|\psi_{1776}(\BFq)|^2$  
displayed in fig.~\ref{fig:stadium-two-C}b)
shows localization along the $\overline{1133}$ orbit.
From this one would expect two major directions in  momentum space.
However, this is not seen in the pictures of
$|\widehat{\psi}_{1776}(\BFp)|^2$ and  $I_{1776}(\varphi)$;
one observes several significant momentum directions,
and at first sight no simple relation to the plot of $|\psi_{1776}(\BFq)|^2$
seems to exists. But there are three peaks in $I_{1776}(\varphi)$ 
close to one of the directions of the triangular orbit, so there is an 
enhanced probability around the orbit in momentum space. 
This seems to be a common phenomenon, when interpreting the pictures 
of $I_n(\varphi)$, one is sometimes mislead by the height of some peaks, 
the relevant information is the area under the curve 
around this peak, which can  
be quite small due to the narrowness of some of the peaks.  

Fig.~\ref{fig:stadium-two-Da} shows two 
eigenfunctions with clear localization
in position space, this time for the stadium billiard
with even--even symmetry. The first is localized
along the unstable $\overline{01}$ orbit running along the symmetry-axis.
This localization is also clearly seen in momentum space,
where $|\widehat{\psi}_{500}(\BFp)|^2$  shows 
strong enhancement in the directions $\varphi=0,\pi$,
corresponding to a strong peak of $I_{500}(\varphi)$
at $\varphi=0$.
In case of the eigenfunction displayed in fig.~\ref{fig:stadium-two-Da}b)
there is also just one distinguished momentum direction.
The orbit $\overline{111333}$ has two momentum directions,
where the one with smaller angle is expected to dominate
due to the longer line segments. This is also
seen in the plot of $I_{1273}(\varphi)$,
where the momentum direction corresponding to the
shorter segments of the orbit do not acquire a high probability.

Fig.~\ref{fig:stadium-two-Db} shows two further examples
of localized eigenfunctions. The first has a nice
pattern in position space, which appear to be associated
to the $\overline{10203020}$ orbit (only one orbit
of the two symmetric partners
is shown in fig.~\ref{fig:stadium-two-Db}a). 
This orbit has just one momentum direction,  however 
in momentum space we observe five prominent momentum directions.
One could associate the three leftmost peaks in $I_{1409}(\varphi)$
to this orbit, the additional peaks might be caused by
some fine structure, which is hardly visible in the plot
of  $|\psi_{1409}(\BFq)|^2$.
The eigenfunction with $n=1993$ displayed in 
fig.~\ref{fig:stadium-two-Db}b) also shows clear localization
in position space along
the $\overline{110552}$ orbit. In momentum space there is one
significant momentum direction at $\varphi\approx 3\pi/8$, 
however this direction does not correspond to the pattern
seen in position space, from which one would expect
an important contribution to $|\widehat{\psi}_{1993}(\BFp)|^2$  
and $I_{1993}(\varphi)$ in the direction
of $\varphi=0.71\ldots\,$. Indeed near to this direction
there is a range of angles 
for which $I_{1993}(\varphi)$ has a number of smaller peaks, which 
together lead to an enhanced probability in this interval 
compared to  other intervals.
This example again illustrates that not just the height of the peaks
is of importance in the interpretation
of the pictures for $I_n(\varphi)$, but
the overall accumulated probability corresponding
to some interval of momentum directions.

\newcommand{\eindreiDbildstadion}[2]{
   \centerline{$n=#1$, odd--odd symmetry}

   \hspace*{-0.75cm}
   \begin{minipage}{9cm}
     \centerline{\PSImagx{wfk_3d_stadion_#1.ps}{7.8cm}}
   \end{minipage}
   \begin{minipage}{9cm}
     \centerline{\PSImagx{ft_3d_stadion_#1.ps}{7.8cm}}
   \end{minipage}


   \hspace*{-0.75cm}
   \begin{minipage}{9cm}
     \centerline{\PSImagxrotated{wfk_stadion_full_#1.ps}{3cm}{-90}} 
   \end{minipage}
   \begin{minipage}{9cm}
     \centerline{\hspace*{0.5cm}\PSImagxrotated{ft_stadion_#1.ps}{7cm}{-90}}
   \end{minipage}


   \hspace*{-0.75cm}
   \begin{minipage}{9cm}
     \centerline{{\PSImagx{stad_ft_vert_#1.ps}{8.5cm}}} 
   \end{minipage}
   \begin{minipage}{9cm}
     \centerline{\PSImagx{ft_vert_rad_stadion_#1.ps}{8.5cm}} 
   \end{minipage}

}

\newcommand{\einbildstadion}[3]{
   {#3) $n=#1$, odd--odd symmetry}

   \centerline{\PSImagxrotated{wfk_stadion_full_#1.ps}{3cm}{-90}} 
 
   \vspace*{0.5cm}

   \centerline{\PSImagxrotated{ft_stadion_#1.ps}{7cm}{-90}}    
 
   \vspace*{0.5cm}
        
   \hspace*{-1cm}
   \PSImagx{stad_ft_vert_#1.ps}{8.5cm} 
   }

\newcommand{\einbildstadionNN}[3]{
   {#3) $n=#1$, even--even symmetry}

   \centerline{\PSImagxrotated{wfk_stadion18NN_full_#1.ps}{3cm}{-90}} 
 
   \vspace*{0.5cm}

   \centerline{\PSImagxrotated{ft_stadion18NN_#1.ps}{7cm}{-90}}  
 
   \vspace*{0.5cm}
           
   \hspace*{-1cm}
   \PSImagx{stad18NN_ft_vert_#1.ps}{8.5cm}   
           
}

\newcommand{\zweibildstadion}[4]{
  \begin{minipage}{9.25cm}
    \einbildstadion{#1}{#2}{a}
  \end{minipage}  
  \begin{minipage}{9.25cm}
    \einbildstadion{#3}{#4}{b}    
  \end{minipage}  
}

\newcommand{\zweibildstadionNN}[4]{
  \begin{minipage}{9.25cm}
    \einbildstadionNN{#1}{#2}{a}
  \end{minipage}  
  \begin{minipage}{9.25cm}
    \einbildstadionNN{#3}{#4}{b}    
  \end{minipage}  
}

\BILD{tbh}
     {
       \eindreiDbildstadion{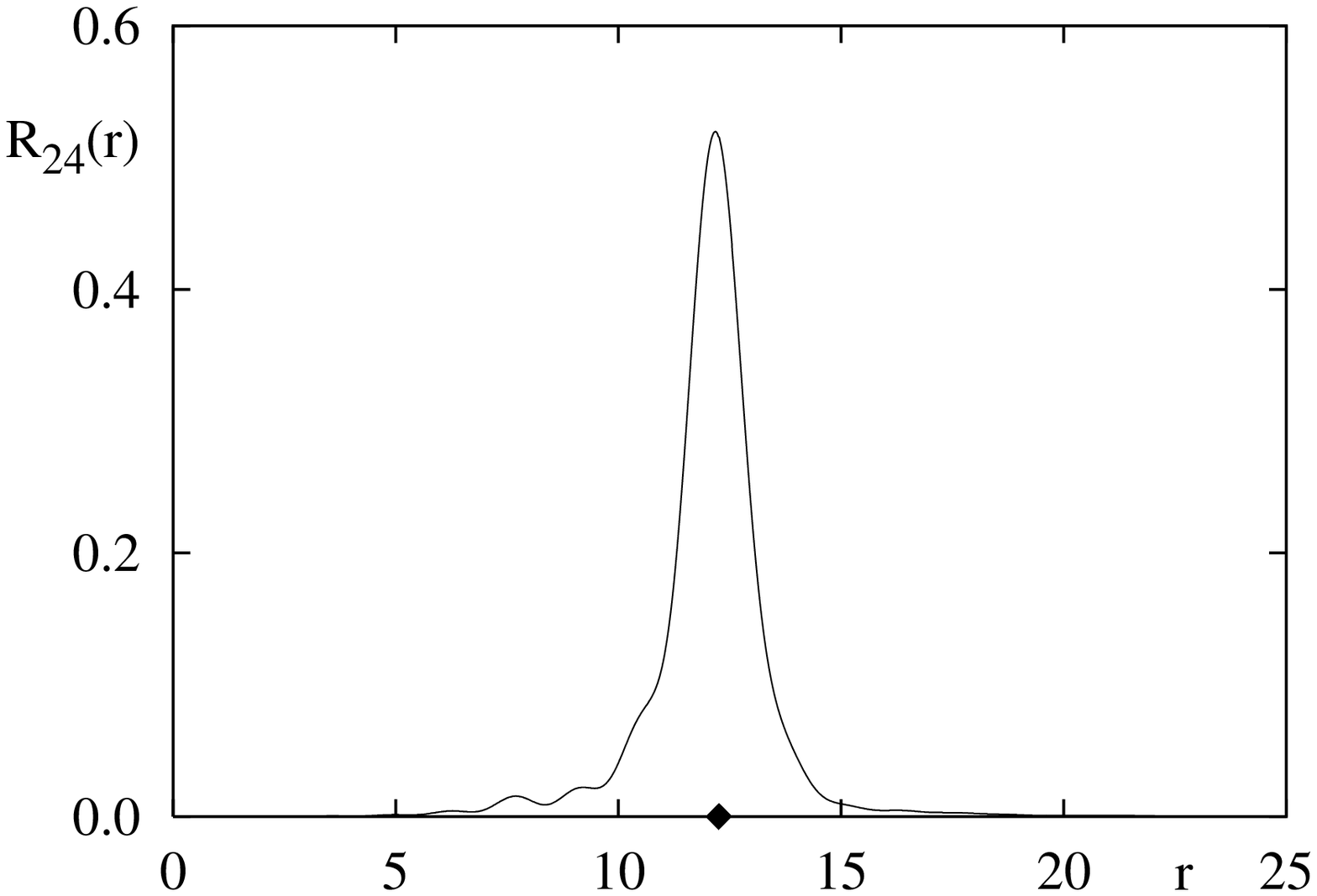}{150.1094}
     }
     {Three--dimensional plots of $|\psi_{24}(\BFx)|^2$, 
     $|\widehat{\psi}_{24}(\BFp)|^2$, their corresponding grey--scale 
     pictures and the plot of the radially integrated momentum
     distribution $I_{24}(\varphi)$ and the angular
     integrated momentum distribution $R_{24}(r)$.
     The momentum distribution $|\widehat{\psi}_{24}(\BFp)|^2$
     is concentrated around the energy shell, which is indicated as 
     the inner circle. This is also clearly
     visible in the plot of $R_{24}(r)$.
     The angular distribution $I_{24}(\varphi)$ does not show 
     any significantly preferred directions and the plot of 
     $|\psi_{24}(\BFx)|^2$ also does not show any prominent patterns.}
     {fig:stadium-ThreeD-A}

\BILD{tbh}
     {
       \eindreiDbildstadion{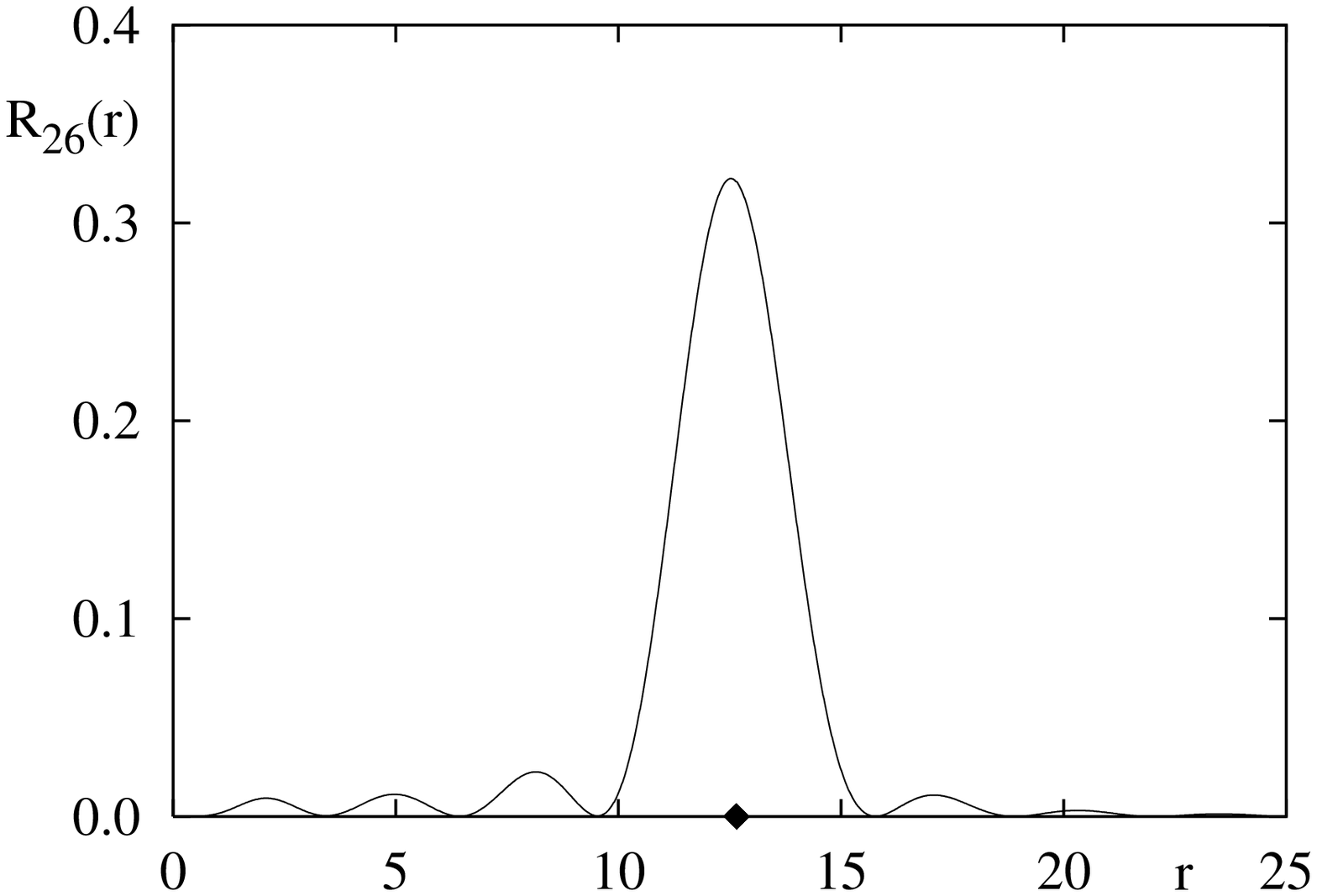}{160.0318}
     }
     {Example of a low--lying bouncing ball mode, for which 
      $|\psi_{26}(\BFx)|^2$ is localized on the rectangular part of the 
      billiard. The plot of the momentum distribution 
      $|\widehat{\psi}_{26}(\BFp)|^2$ shows a strong localization
      in the momentum directions ($p_x\approx 0$, $p_y\approx\pm p_n$)
      of the bouncing ball orbits. This is also clearly seen in the plot
      of  $I_{26}(\varphi)$, which is concentrated near to $\varphi=\pi/2$.
      The corresponding direction 
       $\varphi^{\text{bb}}_{2,8} =\arctan(4 a)$
       is marked by a triangle.
      In the plot of $R_{26}(r)$ one observes some
      additional oscillations in comparison to $R_{24}(r)$ 
      in fig.~\ref{fig:stadium-ThreeD-A}.
      }
     {fig:stadium-ThreeD-B}

\BILD{tbh}
     {

     \zweibildstadion{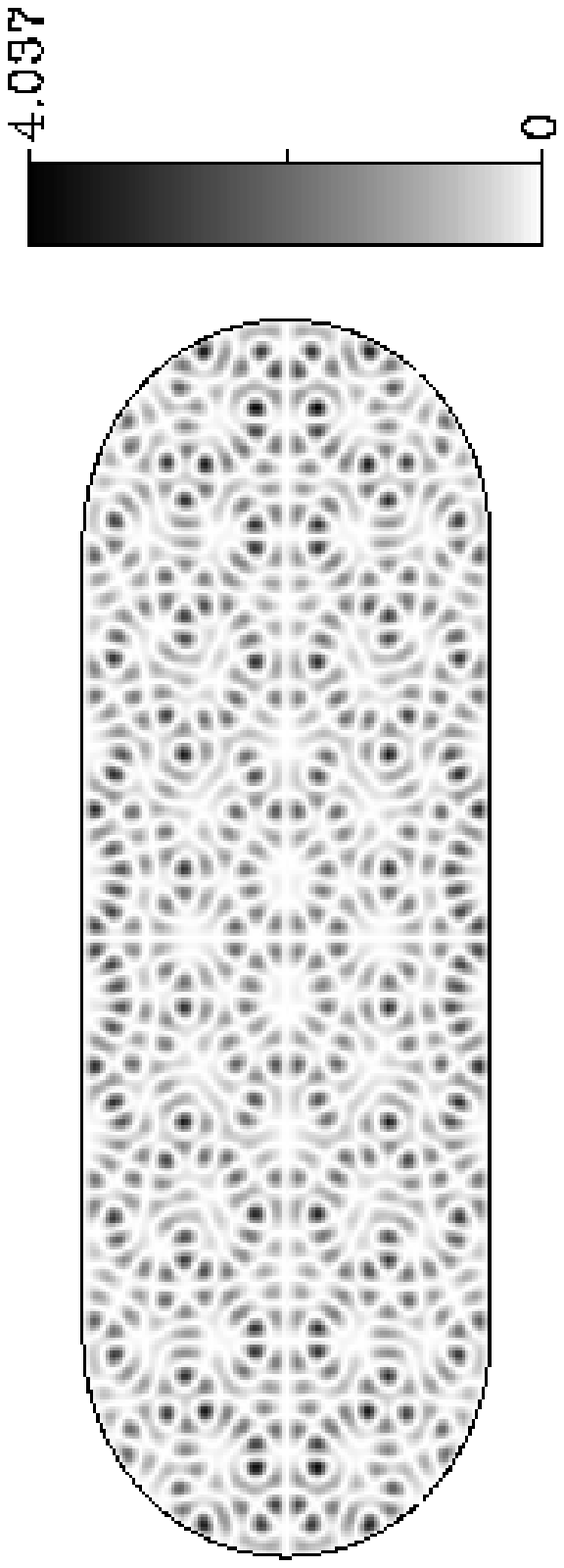}{.}{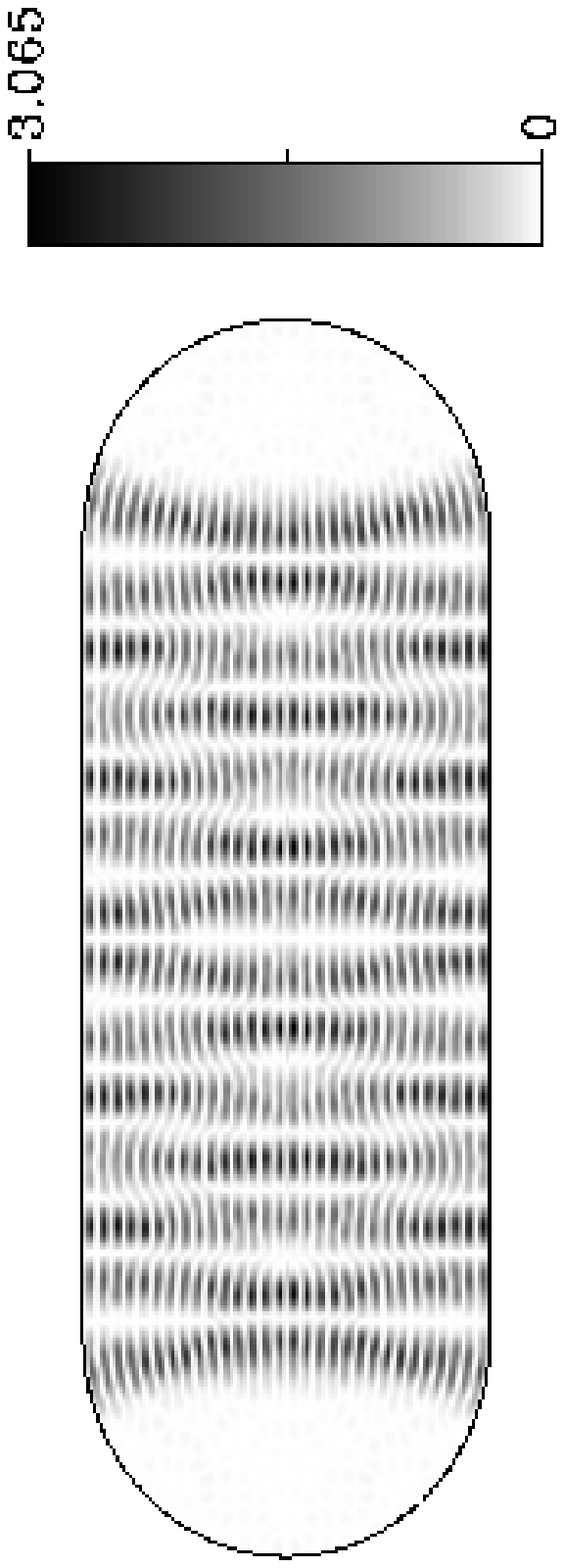}{.}

     }
     {In a) an example of an eigenfunction is shown which
      appears to be completely non--localized in position space. However,
      in momentum space $|\widehat{\psi}_{263}(\BFp)|^2$
      and $I_{263}(\varphi)$ show clear localization in three major
      momentum directions.
      The second eigenfunction is an example of a higher lying 
      bouncing ball mode with $14$ modes in $x$ direction
      and $30$ modes in $y$ direction. Consequently the corresponding
      momentum distributions show localization near to the $\varphi=\pi/2$
      direction.
      In the plot of $I_{455}(\varphi)$ the triangle 
      marks the direction $\varphi^{\text{bb}}_{14,30} =\arctan(30/14\; a)$.}
     {fig:stadium-two-B}

\BILD{tbh}
     {
     \zweibildstadion{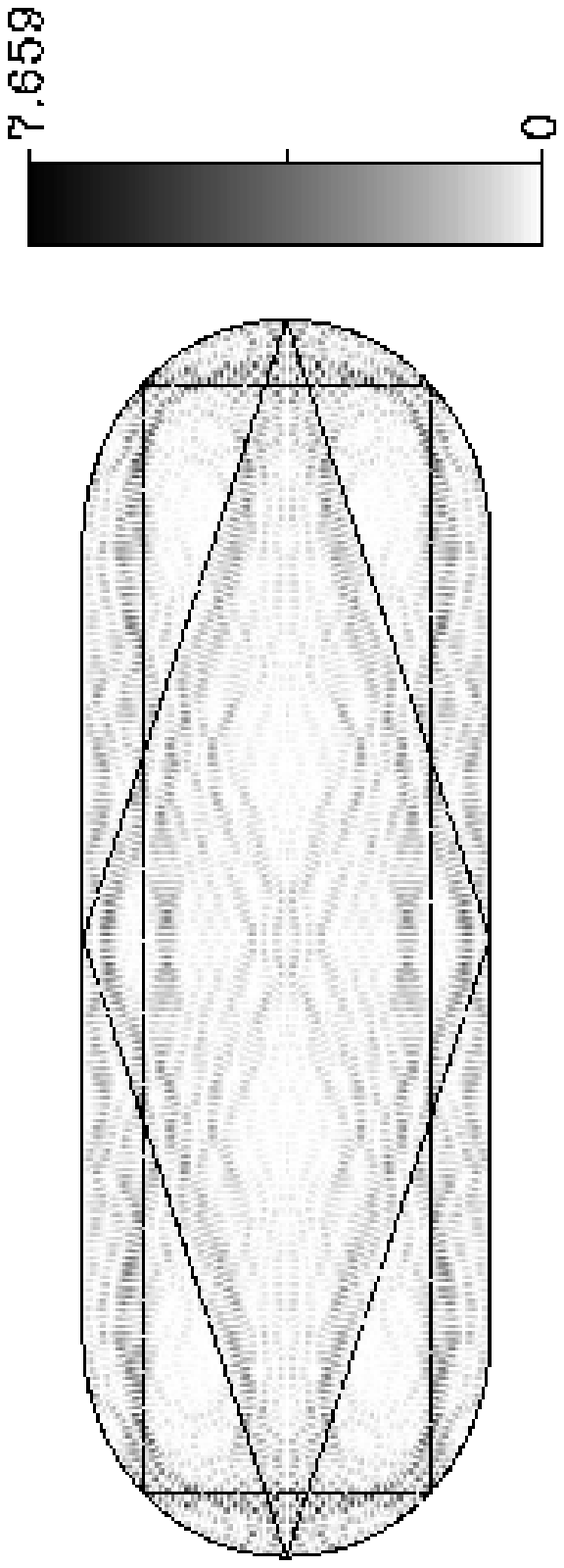}{.}{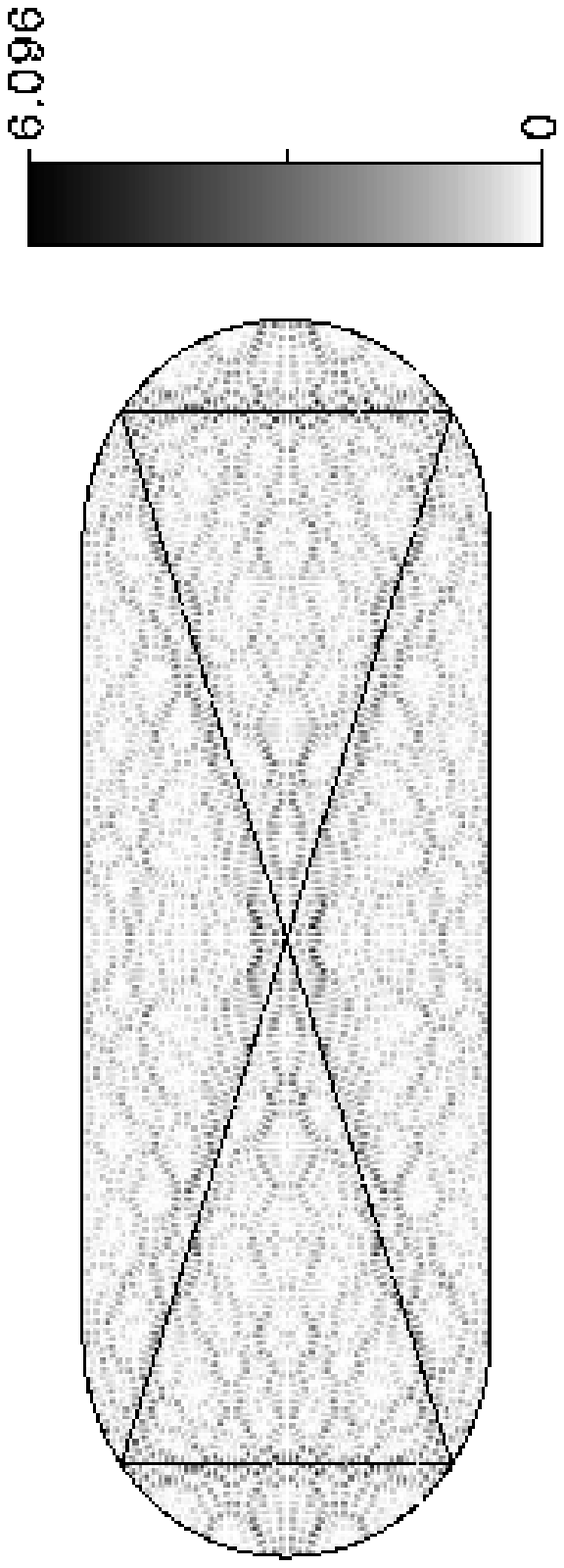}{.}
     }
     {In a) the  eigenfunction shows localization 
      along the $\overline{1155}$ orbit of rectangular shape and 
      also the $\overline{0123}$ orbit in the stadium.
      In the plot of $I_{1771}(\varphi)$ the corresponding
      momentum directions are marked by 
      full triangles ($\overline{1155}$ orbit) 
      and open triangles ($\overline{0123}$ orbit).
      The  eigenfunction in b) shows scarring along the $\overline{1144}$ 
      orbit, which is clearly seen in the plot of 
      $|\psi_{1776}(\BFx)|^2$. However, in this case
      the scarring is not that clearly reflected in the pictures
      of $|\widehat{\psi}_{1776}(\BFp)|^2$
      and $I_{1776}(\varphi)$, see the text for a discussion.
     }
     {fig:stadium-two-C}

\BILD{tbh}
     {
      \zweibildstadionNN{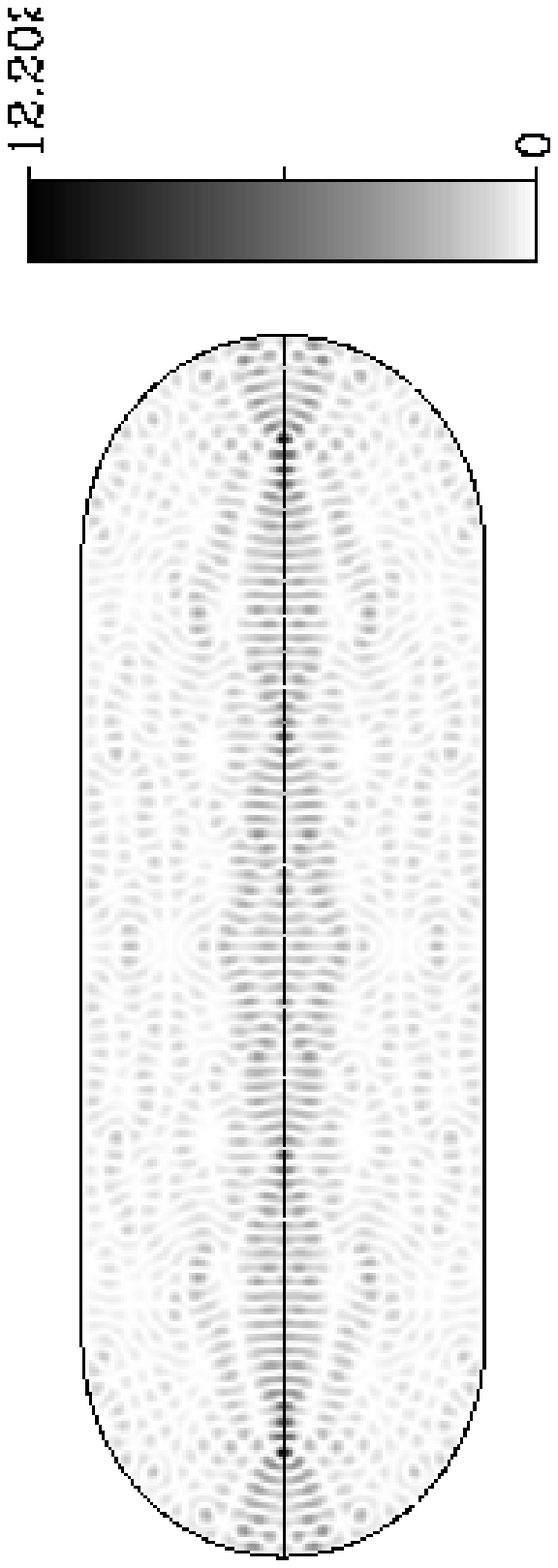}{.}{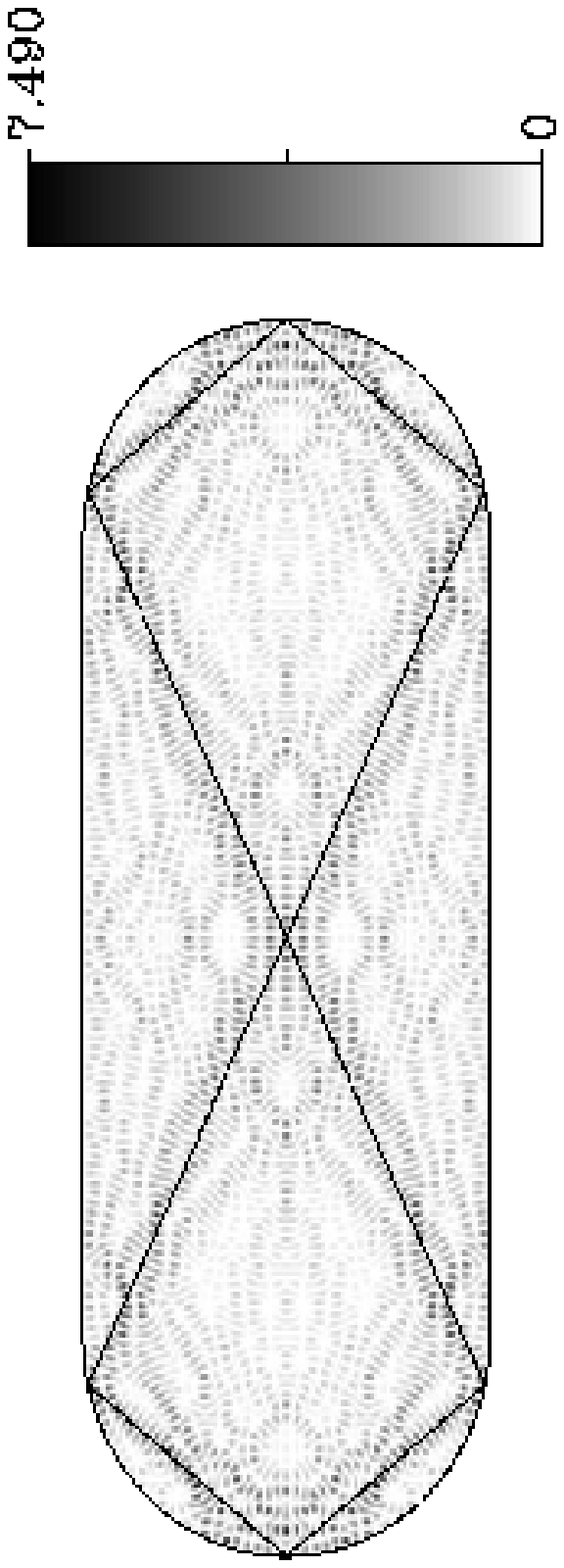}{.}
      
     }
     {For $n=500$ one has a nice example of an eigenfunction localized around
      the shortest unstable orbit, which runs along the symmetry axis.
      This is also seen in the plots of $|\widehat{\psi}_{500}(\BFp)|^2$
      and $I_{500}(\varphi)$, which show a strong enhancement
      in the $\varphi=0,\pi$ directions.
      The  eigenfunction in b) shows localization around the 
      $\overline{111333}$ orbit.}
     {fig:stadium-two-Da}

\BILD{tbh}
     {
      \zweibildstadionNN{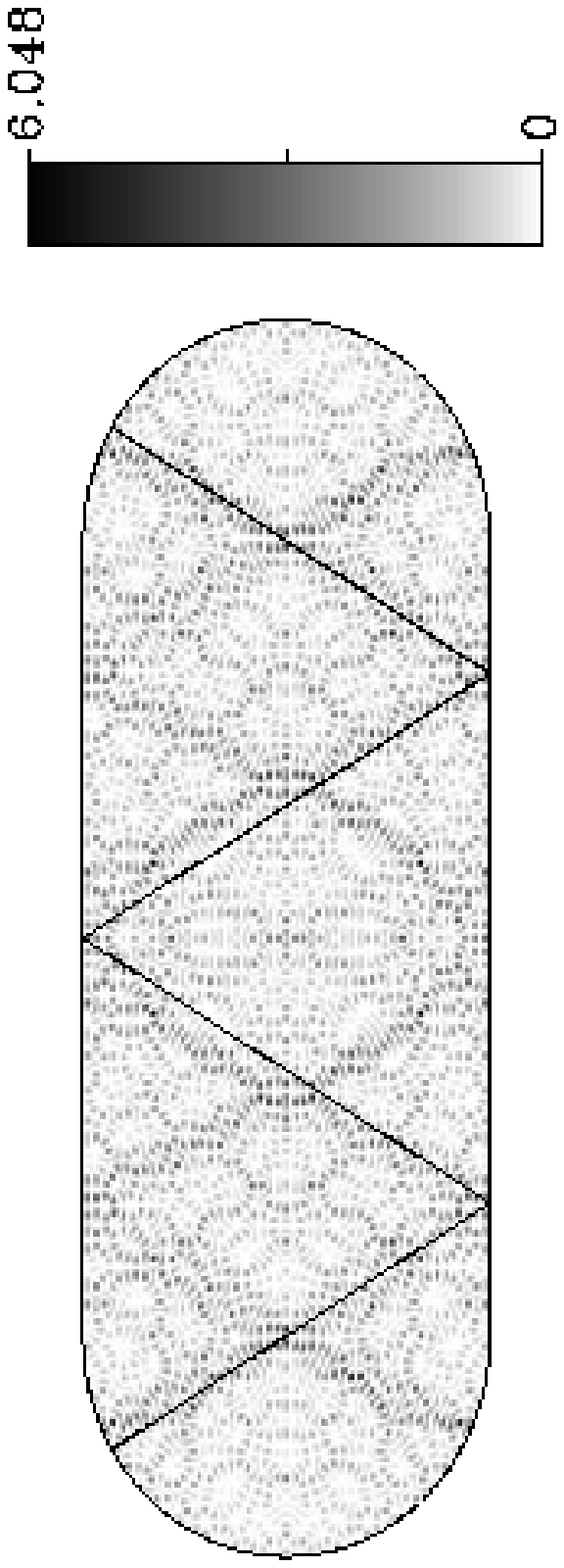}{.}{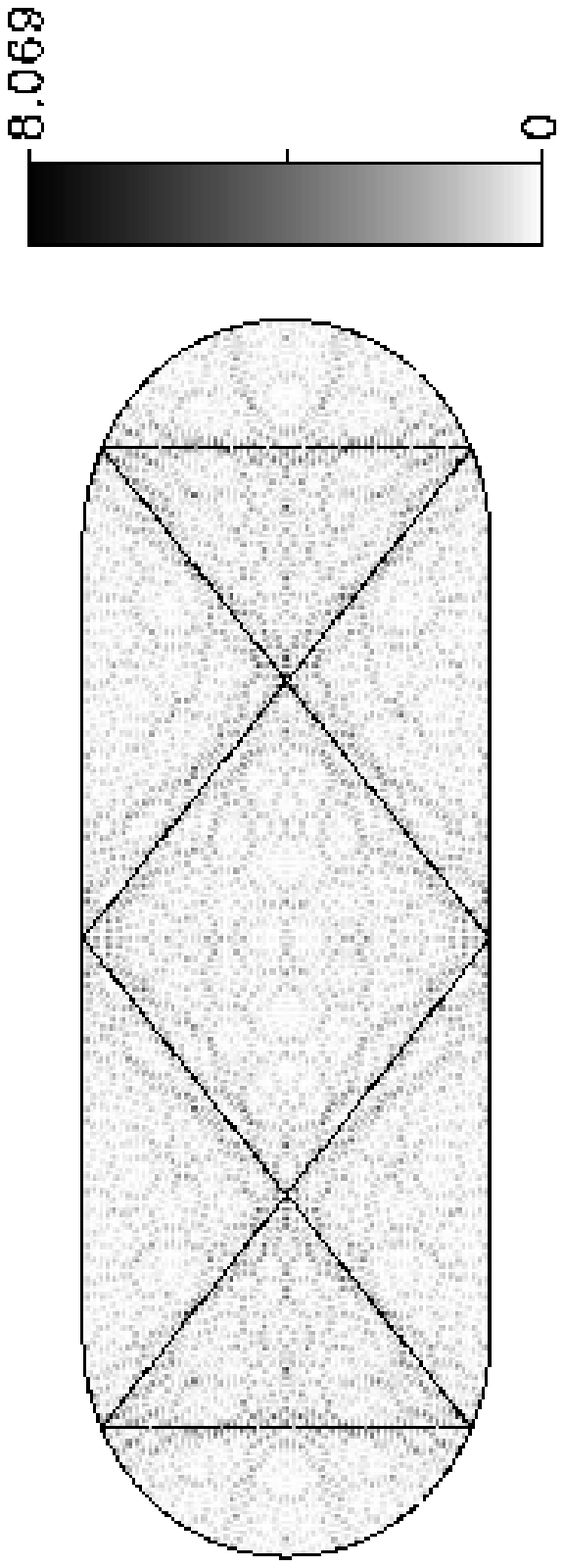}{.}      
     }
     {The eigenfunction with index $n=1409$ of even--even symmetry
      is localized along the $\overline{10203020}$ 
      orbit (only one of the two symmetry--related
      partners is shown). Surprisingly the plots of 
      $|\widehat{\psi}_{1409}(\BFp)|^2$ and $I_{1409}(\varphi)$
      show five peaks instead of just one, see the text for 
      further discussion.
      The  eigenfunction in b) is localized in position space along
      the  $\overline{1105552}$ orbit. This orbit has two different momentum
      directions (in the intervall $\varphi\in[0,\pi/2]$),
      still there is an additional high peak visible for 
      $I_{1993}(\varphi)$.  This illustrates
      that for $I_n(\varphi)$ not just the height of the peak
      is relevant, but the total area below a peak, which accounts
      for a high probability in the corresponding direction.}
     {fig:stadium-two-Db}

\BILD{tbh}
     {
      \zweibildstadion{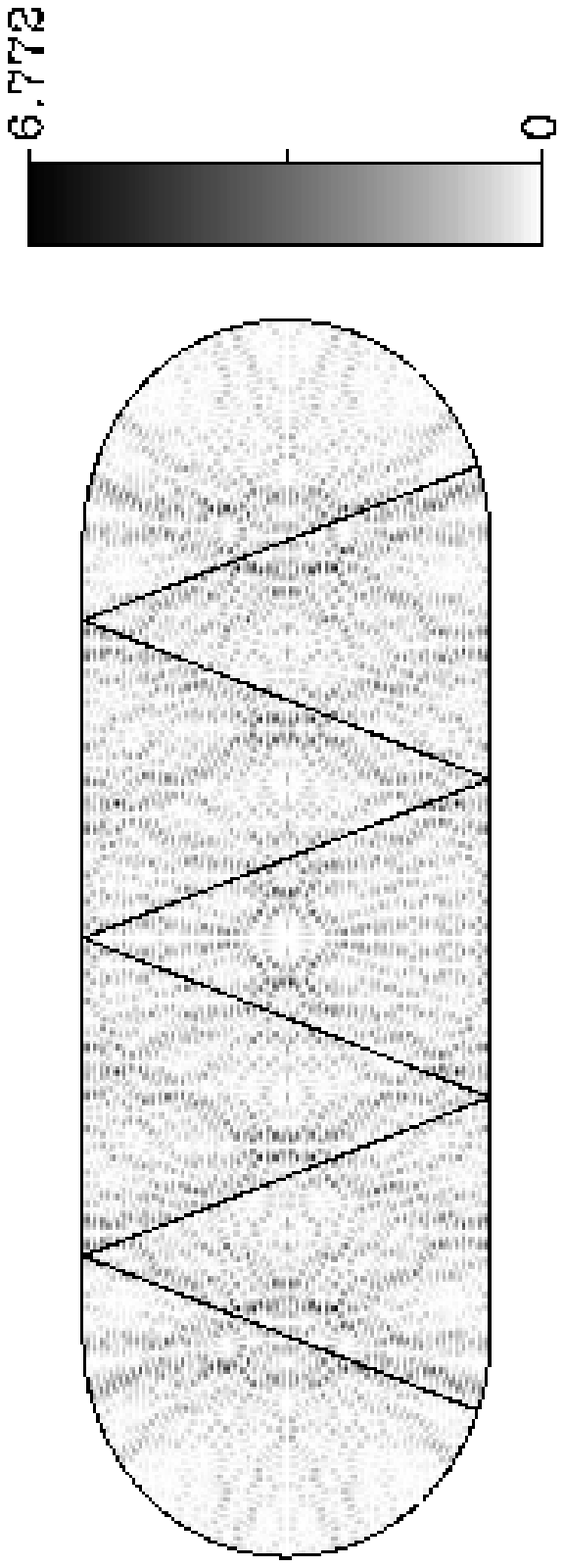}{.}{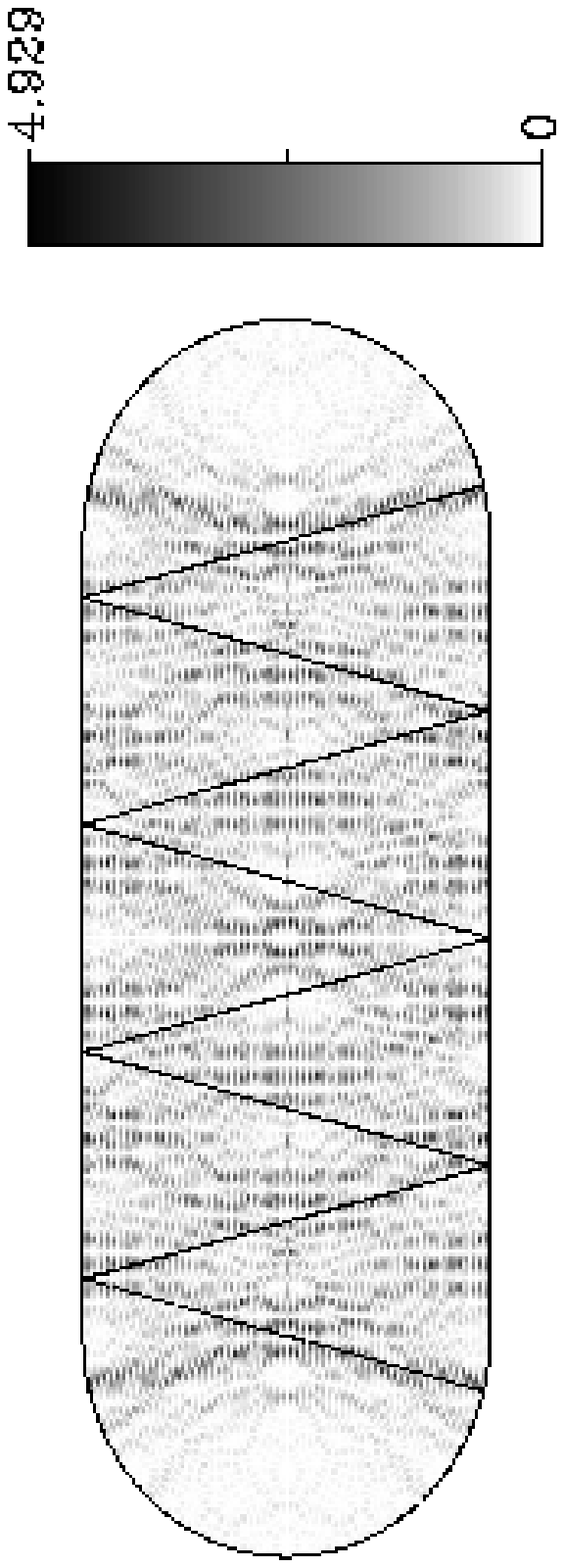}{.}     
     }
     {Two examples of eigenfunctions which are localized on the 
      rectangular part
      of the stadium billiard, which are not bouncing ball--modes. 
      These eigenfunctions may be understood in terms of 
      the plotted orbits, although in momentum space 
      there are additional prominent momentum directions.
      This might indicate that both orbits contribute to each
      eigenfunction, see the text for further discussion.}
     {fig:stadium-two-E}

\afterpage{\clearpage}

In \cite{BaeSchSti98} the rate of quantum ergodicity
has been studied both in position and in momentum space.
For the stadium billiard it was expected
that if the bouncing ball modes
are the dominating (in the sense of having
the strongest increase in the counting function) 
subsequence of localized eigenfunctions,
that then the rate of quantum ergodicity should
obey 
$S_1(E,A) := \frac{1}{N(E)}  \sum_{E_n \le E }| \langle \psi_n, A \psi_n\rangle
  - \overline{\sigma(A)}|
 \sim E^{-1/4}$.
However, it turned out that the rate
is much slower \cite{BaeSchSti98}, at least in the considered
energy range covering the first $6\,000$ eigenfunctions.
This was attributed to a considerable
number of eigenfunctions, which show localization in position
space in the rectangular part of the billiard, without 
being bouncing ball modes, see fig.~11 of \cite{BaeSchSti98}.
Two examples are shown in fig.~\ref{fig:stadium-two-E}. 
One sees that they are also localized in momentum space 
near the direction $\varphi=\pi/2$, like the bouncing ball modes, 
but have more peaks. These states seem to correspond 
to the set of periodic orbits which are bouncing $m$ times 
between the parallel walls before they are reflected into 
themselves in the circular part. Two of them, the 
$\overline{202023202024}$ and the $\overline{2020202320202024}$ orbit,
are shown in the density plots of $|\psi_{1652}(\BFx)|^2$ and 
$|\psi_{1872}(\BFx)|^2$.
Since the lengths of these orbits are close to being rationally 
dependent one can speculate, 
that the naive Bohr-Sommerfeld quantization leads to 
two eigenvalue-sequences which have some very close pairs of 
eigenvalues, for which the corresponding eigenfunction should be scarred by 
both orbits. The plots of $I_{1652}(\varphi)$ and 
$I_{1872}(\varphi)$ indicate that this happens, because both 
have strong peaks at the same directions corresponding to the two 
orbits.

The cardioid billiard 
possesses no parallel walls and therefore no
bouncing ball modes; in this sense it is more generic than the stadium 
billiard. Figs.\ \ref{fig:cardid-ThreeD-A}--\ref{fig:cardid-ThreeD-B} 
show for the cardioid billiard three--dimensional plots of 
$|\psi_n(\BFx)|^2$, 
$|\widehat{\psi}_n(\BFp)|^2$
and the corresponding grey-scale plots. Furthermore 
$I_n(\varphi)$ and $R_n(r)$ are shown.
The eigenstate of odd symmetry with $n=24$ shows localization
along the unstable orbit $\overline{AB}$, wich is shown in the plot too. 
This localization is also seen in the plots
of $|\widehat{\psi}_{24}(\BFp)|^2$ and $I_{24}(\varphi)$.
In the plot of $I_{24}(\varphi)$ the corresponding direction
of the orbit is marked by a full triangle at $\varphi=\pi/2$.
The plot of $R_{24}(r)$ shows the expected localization
(plus some oscillations) around the energy shell $r=|\BFp_n|$,
indicated by the rhombus. In the plots 
of the Fourier transforms the energy shell is marked
by the full inner circle.
The eigenstate displayed in fig.~\ref{fig:cardid-ThreeD-B}
does not show any particular localization both in 
position and in momentum space.

In fig.~\ref{fig:kardi-two-A} the first example shows clear
localization in position space along the orbit with
code $\overline{AAABBB}$.
In momentum space the vertical direction ($\varphi=\pi/2$) shows
an enhanced probability, whereas the second direction, $\varphi=0.62\ldots$,
is not as prominent as one might expect from the picture
of the eigenfunction in position space.
The second eigenfunction shown in fig.~\ref{fig:kardi-two-A}
is an example of a non--localized eigenfunction, which is also
nicely seen in the plots of the momentum distributions,
which do not show any preferred momentum direction.

Fig.~\ref{fig:kardi-two-B} shows two examples of eigenfunctions
localized along the $\overline{AB}$ orbit,
which is clearly seen in the plots of $|\psi_{1817}(\BFx)|^2$ 
and $|\psi_{2605}(\BFx)|^2$. 
The latter eigenfunction is the 
one which was found in \cite{AurBaeSchTag98:p}
to have the largest maximum norm $||\psi_n||_\infty$
among the first 6000 eigenfunctions of odd symmetry.
Also the corresponding momentum distributions reveal
that the direction $\varphi=\pi/2$ stands out.
However, for the eigenfunction with index $n=2605$,
there are more oscillations visible in the plot of $I_n(\varphi)$,
but still with a high probability near $\varphi=\pi/2$.

\newcommand{\eindreiDbildkardioide}[2]{
   \centerline{$n=#1$, odd symmetry}

   \hspace*{-0.75cm}
   \begin{minipage}{9cm}
     \centerline{\PSImagx{wfk_3d_kardioide_#1.ps}{8.3cm}}
   \end{minipage}
   \begin{minipage}{9cm}
     \centerline{\PSImagx{ft_3d_kardioide_#1.ps}{7.8cm}}
   \end{minipage}


   \hspace*{-0.75cm}
   \begin{minipage}{9cm}
     \centerline{\PSImagxrotated{wfk_kardioide_d_full_#1.ps}{6.0cm}{-90}}
   \end{minipage}
   \begin{minipage}{9cm}
     \centerline{\hspace*{0.5cm}\PSImagxrotated{ft_kardioide_#1.ps}{7cm}{-90}}
   \end{minipage}


   \hspace*{-0.75cm}
   \begin{minipage}{9cm}
     \centerline{{\PSImagx{kardioide_d__ft_vert_#1.ps}{8.5cm}}}
   \end{minipage}
   \begin{minipage}{9cm}
     \centerline{\PSImagx{ft_vert_rad_kardioide_#1.ps}{8.5cm}} 
   \end{minipage}

}

\newcommand{\einbildkardi}[3]{
   {#3)  $n=#1$, odd symmetry}

   \centerline{\PSImagxrotated{wfk_kardioide_d_full_#1.ps}{6cm}{-90}} 
 
   \vspace*{0.5cm}

   \centerline{\PSImagxrotated{ft_kardioide_#1.ps}{7cm}{-90}}   
   \vspace*{0.5cm}

   \hspace*{-0.5cm}
   \PSImagx{kardioide_d__ft_vert_#1.ps}{8.5cm} 
}

\newcommand{\zweibildkardi}[4]{
  \hspace*{-0.5cm}
  \begin{minipage}{9.25cm}
    \einbildkardi{#1}{#2}{a}
  \end{minipage}  
  \begin{minipage}{9.25cm}
    \einbildkardi{#3}{#4}{b} 
  \end{minipage}  
 
}

\BILD{tbh}
     {
       \eindreiDbildkardioide{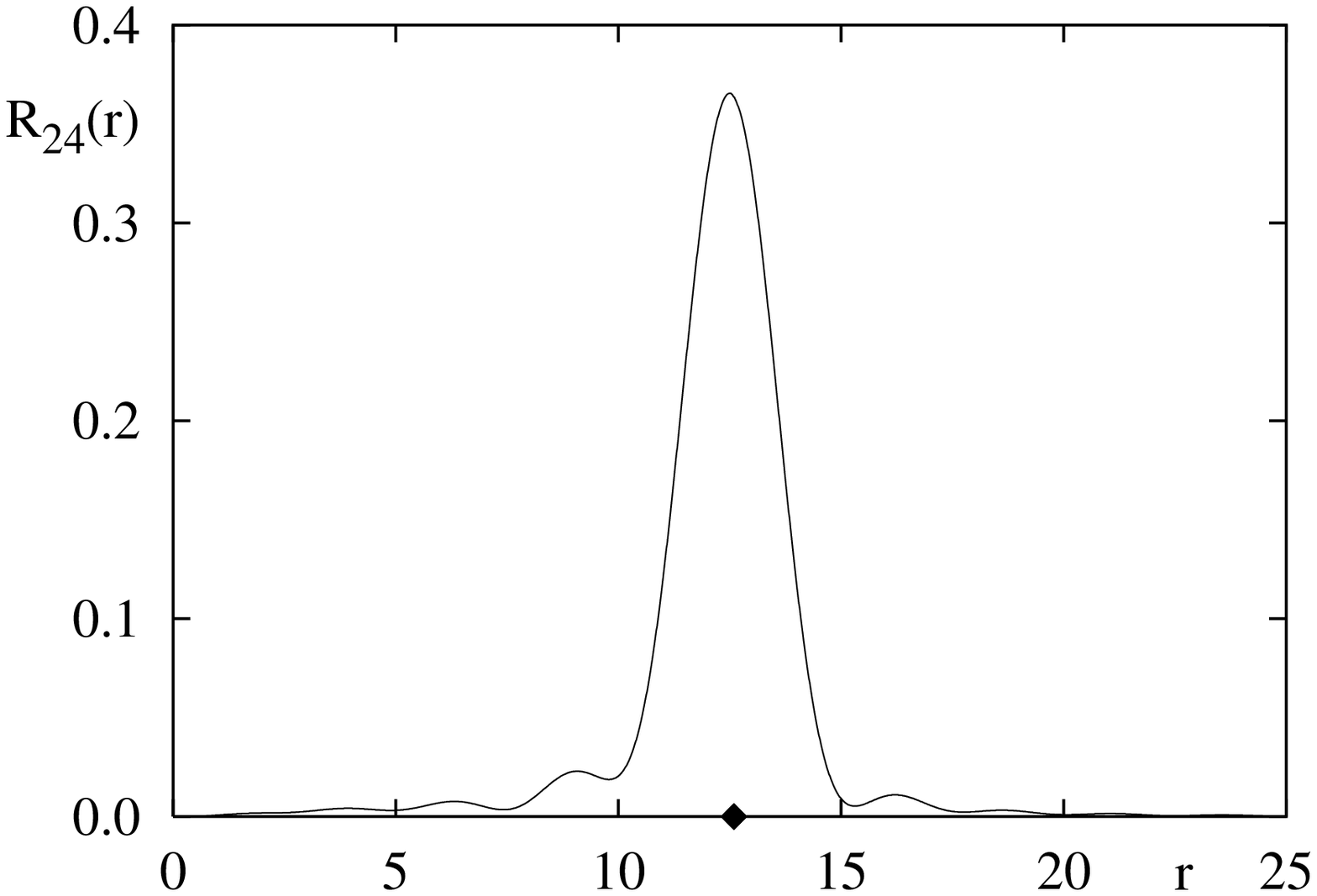}{.}
     }
     {Three--dimensional plots of $|\psi_{24}(\BFx)|^2$, 
     $|\widehat{\psi}_{24}(\BFp)|^2$, their corresponding grey--scale 
     pictures and the plot of the radially integrated momentum
     distribution $I_{24}(\varphi)$ and the angular
     integrated momentum distribution $R_{24}(r)$.
     The momentum distribution $|\widehat{\psi}_{24}(\BFp)|^2$
     is concentrated around the energy shell, which is indicated by the 
     inner circle. This is also clearly
     visible in the plot of $R_{24}(r)$.
     Furthermore this state is to some extent localized along the 
     $\overline{AB}$ orbit, leading to an enhancement of
     $|\widehat{\psi}_{24}(\BFp)|^2$ near to $\varphi=\pi/2,3\pi/2$,
     also seen in the plot of  $I_{24}(\varphi)$ near to the momentum
     direction $\varphi=\pi/2$ (marked by a triangle).
 }
     {fig:cardid-ThreeD-A}

\BILD{tbh}
     {
      \eindreiDbildkardioide{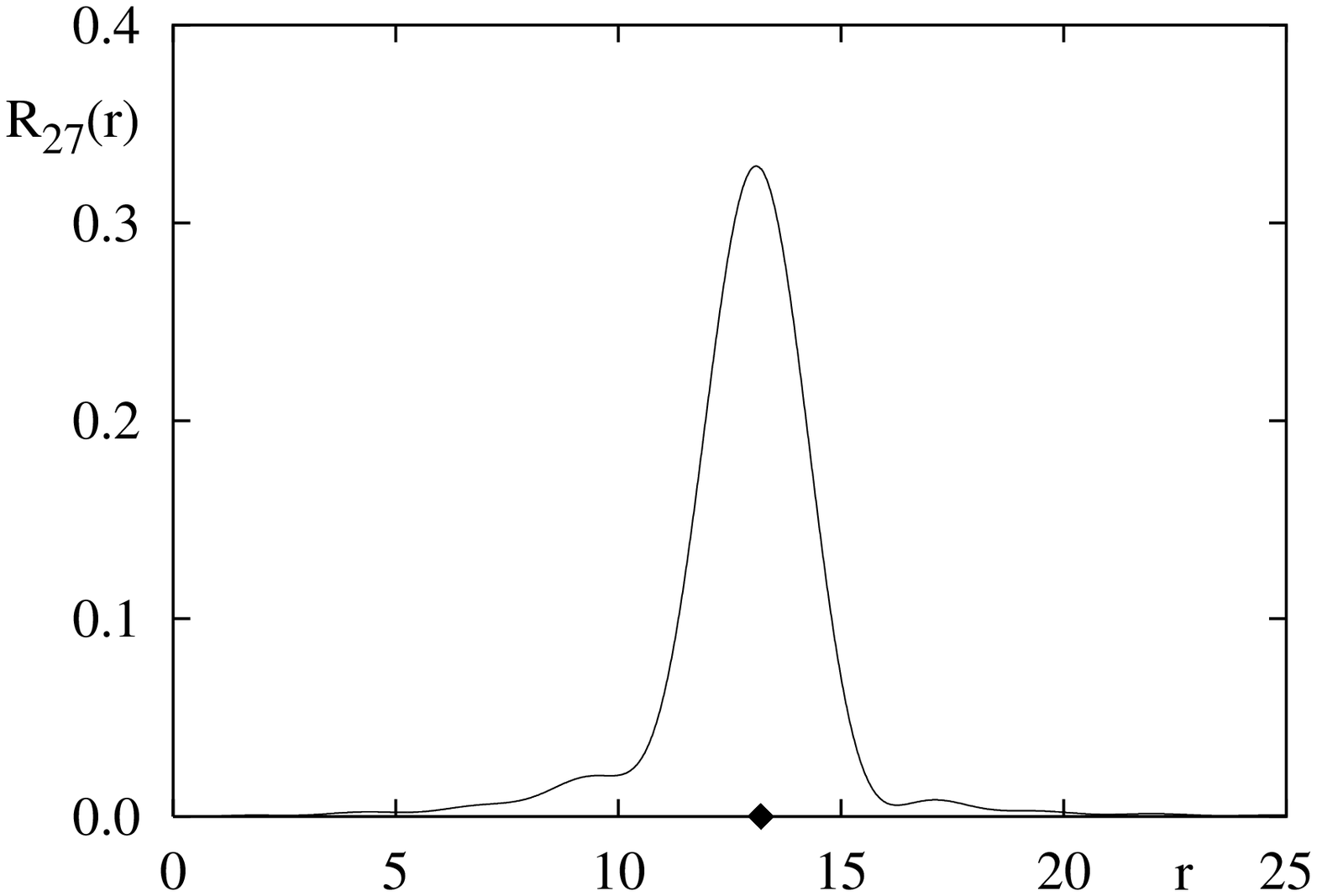}{.}
     }
     {Same as in the previous figure but for $n=27$. 
     In this case there is 
     no prominent localization neither in position nor in momentum space.}
     {fig:cardid-ThreeD-B}

\BILD{tbh}
     {
     \zweibildkardi{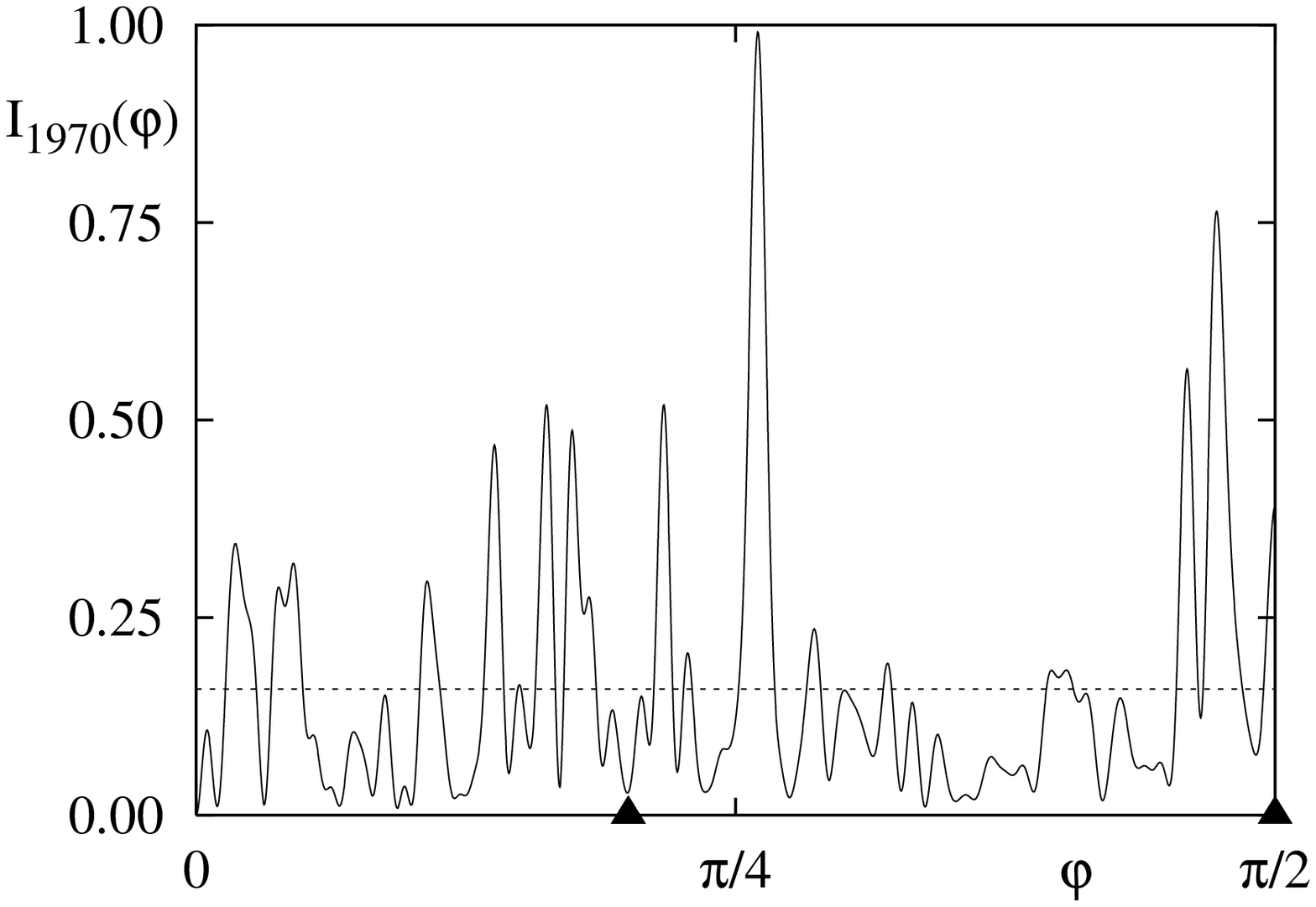}{}{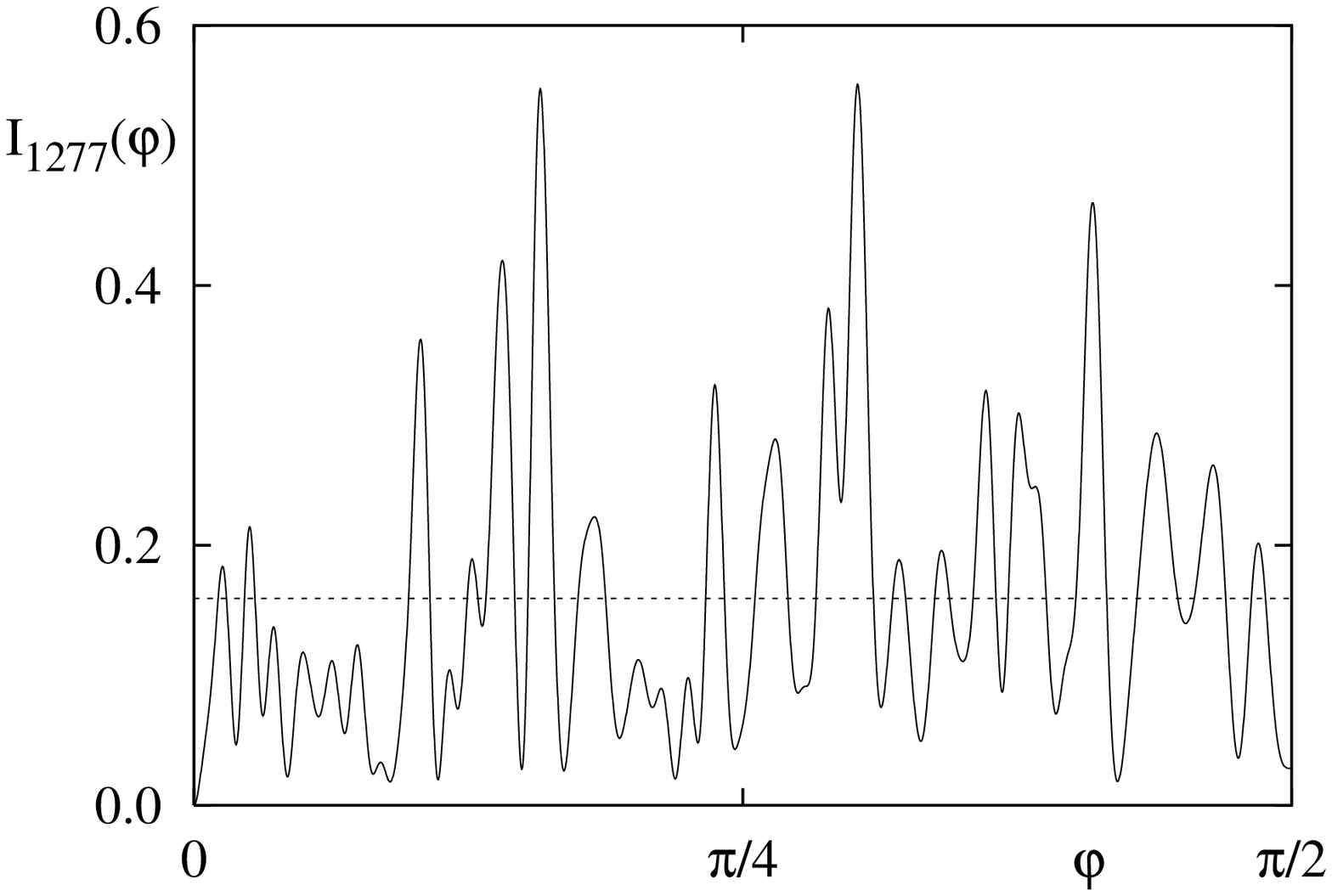}{}
     }
     {The  eigenfunction in a) shows localization along the $\overline{AAABBB}$
      orbit. There is some enhancement in the corresponding
      momentum directions, in particular near $\varphi=\pi/2$, and less
      significantly for the second direction near $\pi/4$.
      The  eigenfunction in b) is an example which appears to be
      quite delocalized both in position and in momentum space. The pictures 
      look like those  expected, according to the quantum ergodicity theorem, 
      for a typical eigenfunction . }
     {fig:kardi-two-A}

\BILD{tbh}
     {
     \zweibildkardi{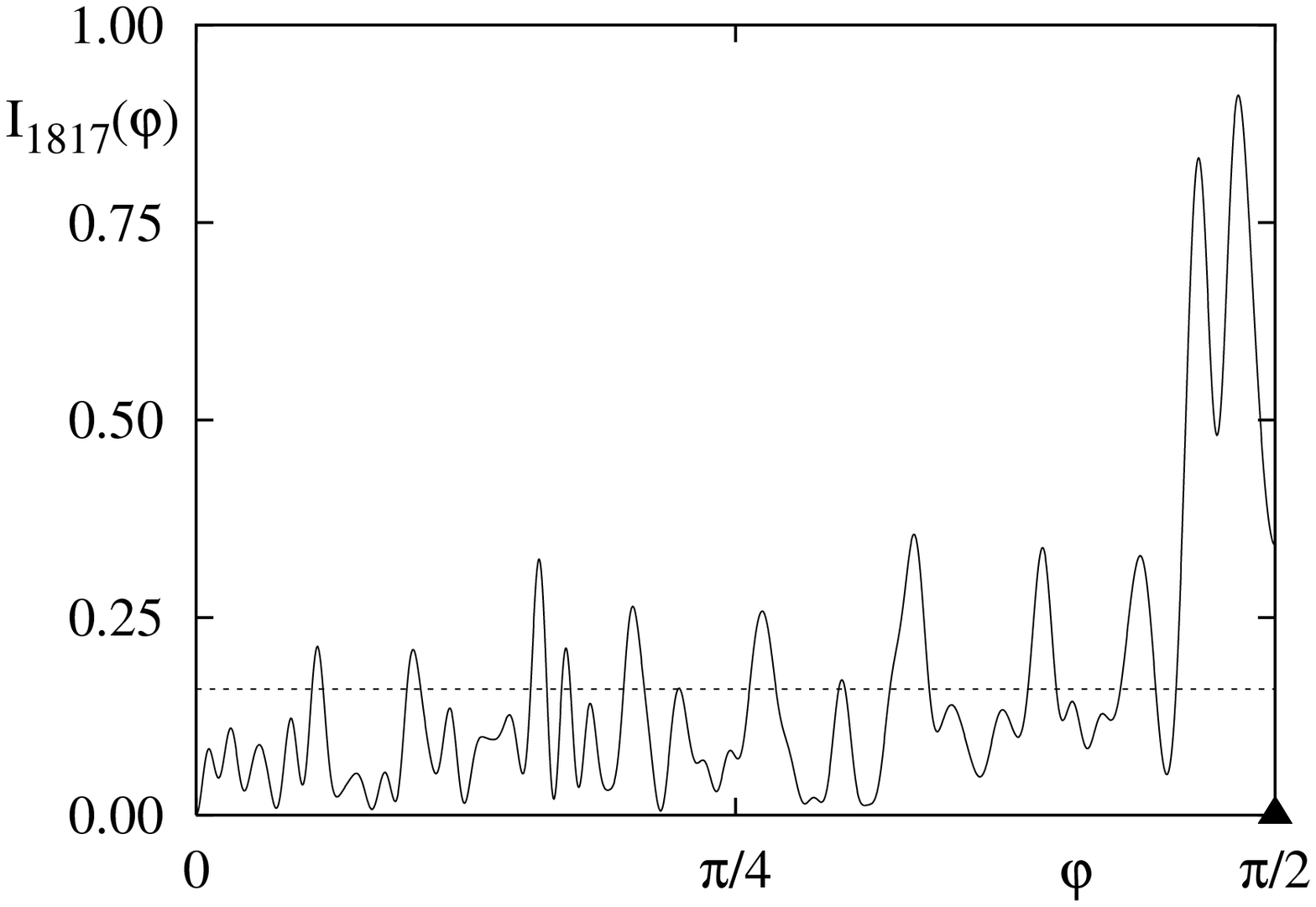}{}{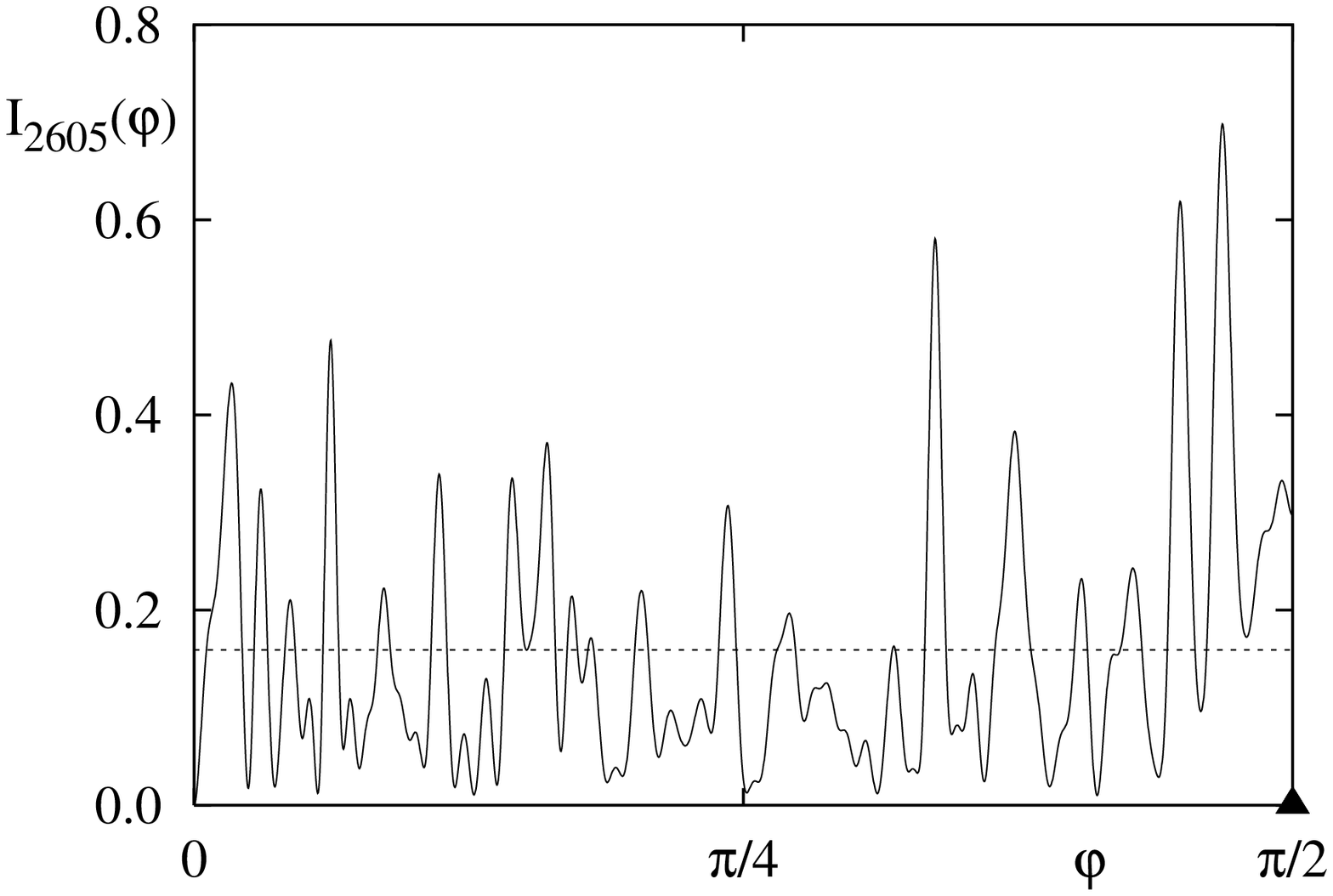}{} 
     }
     {Two examples of eigenfunctions which are localized along the 
     $\overline{AB}$ orbit. In a) the localization appears to be more 
     pronounced than in b). In particular the enhancement 
     near to $\varphi=\pi/2$
     is much less pronounced for $I_{2605}(\varphi)$ 
     in comparison to $I_{1817}(\varphi)$.}
     {fig:kardi-two-B}

\BILD{tbh}
     {
     \zweibildkardi{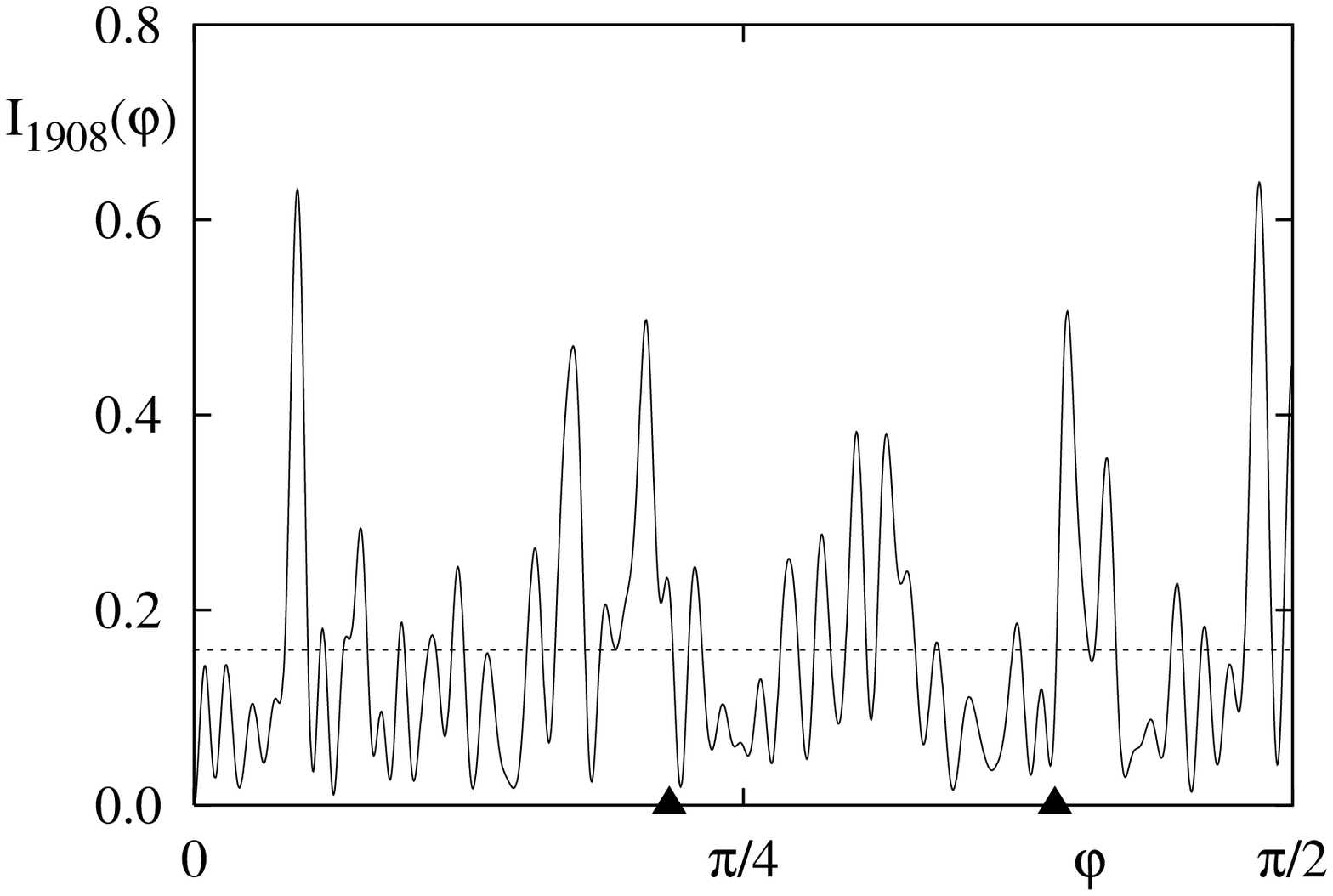}{}{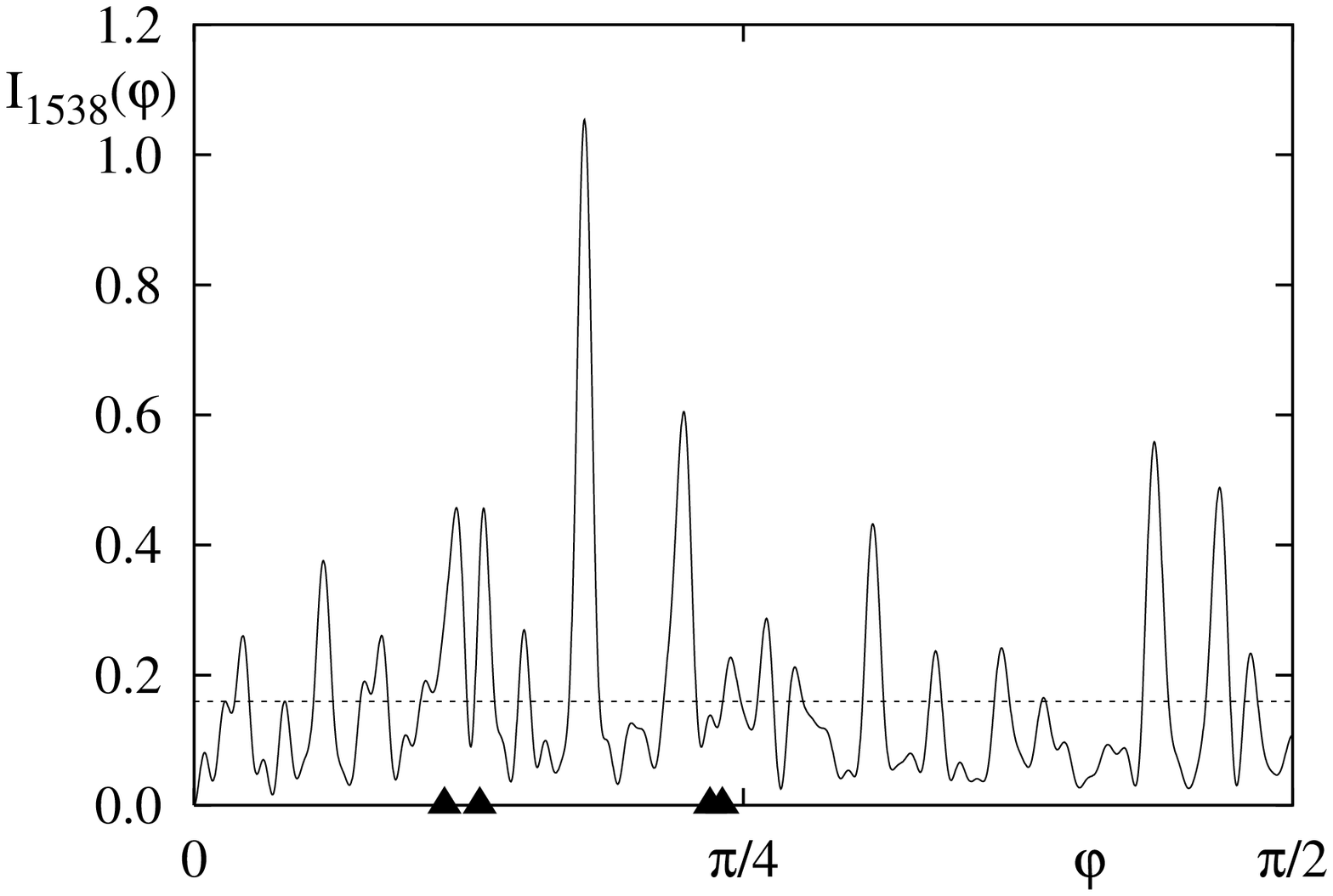}{}
     }
     {The eigenfunction displayed in a) 
      shows localization along the cusp orbit $CAAC$, 
      and thus is an example of a {\it diffractive scar}. 
      The association of the  corresponding momentum directions
      seems to be ambiguous.
      The  second example  shows an eigenfunction, which is mainly
      localized along the boundary. The family of orbits with 
       code $\overline{A^nBABB}$ appears to describe the observed
       pattern quite well. In the plot of $I_{1538}(\varphi)$ only
      the  momentum directions of the part of the orbit running
      inside the billiard are marked by triangles. }
     {fig:kardi-two-C}

\BILD{tbh}
     {
     \zweibildkardi{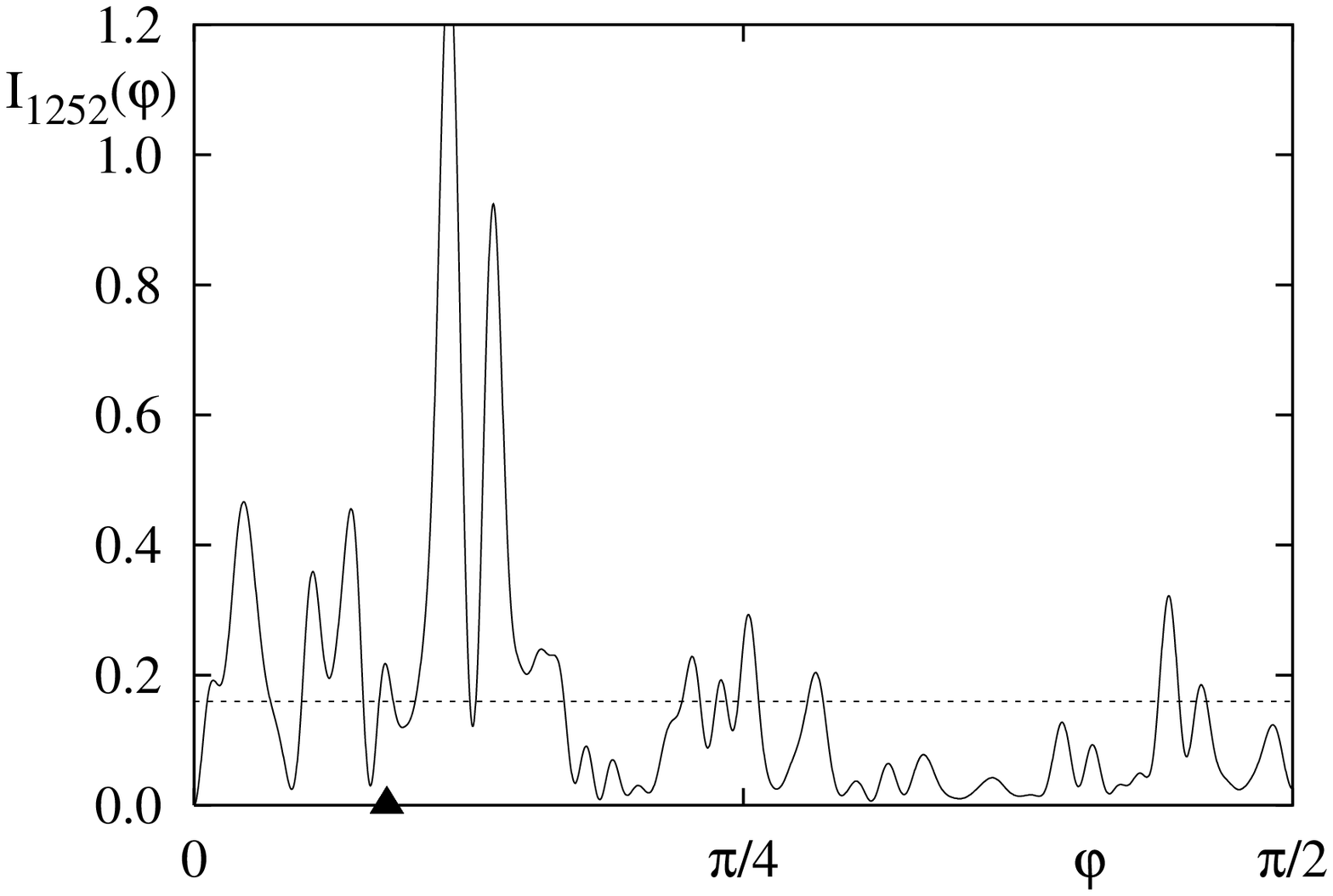}{}{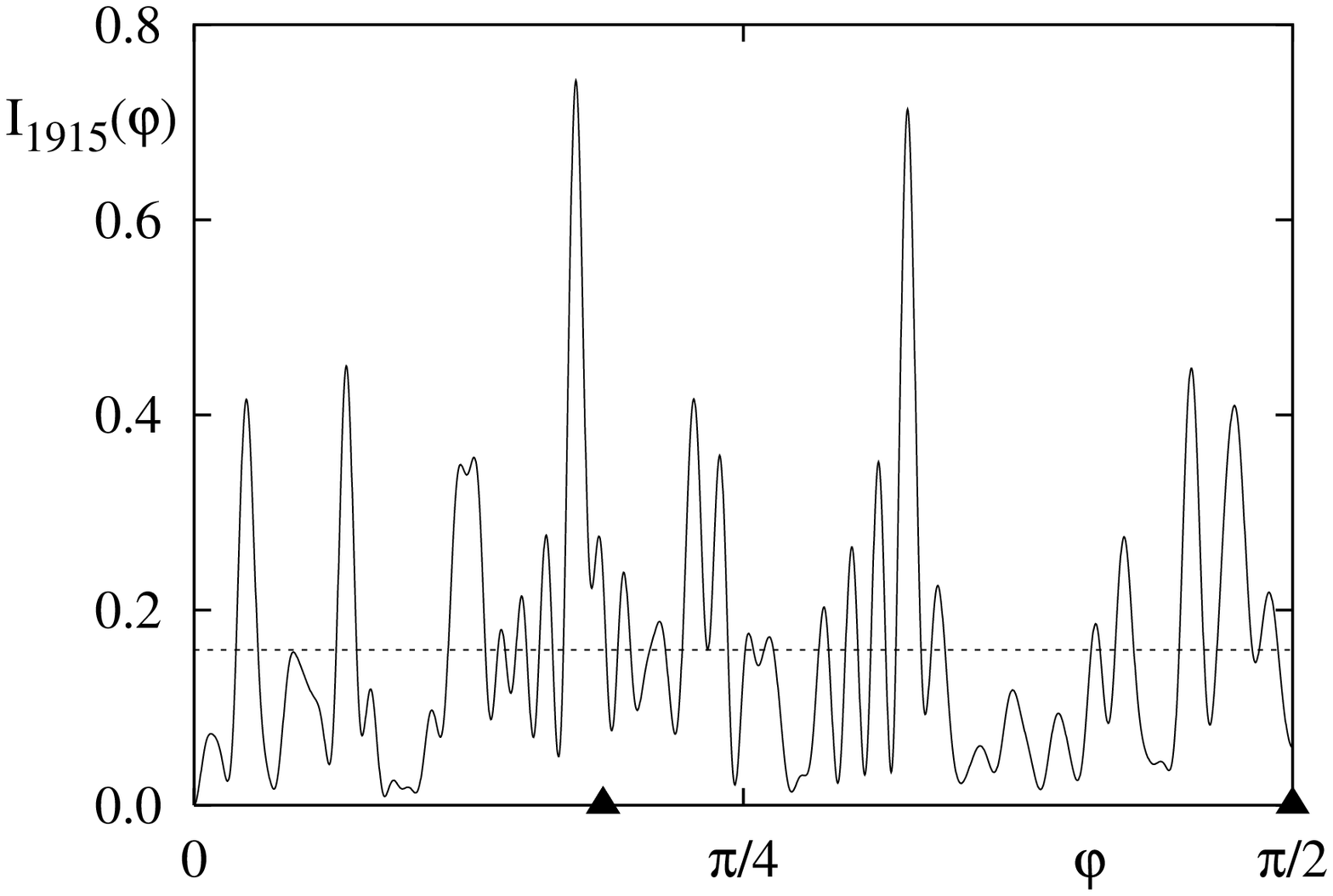}{} 
     }
     {In a) a nice example of an eigenfunction showing scarring
      along the $\overline{AABB}$ orbit is shown. Near to the symmetry
      axis the eigenfunction density is spread out. 
      In momentum space there is a clear enhancement, which is
      close to (although not as close as one might expect)
      the momentum direction of the orbit.
      The  eigenfunction displayed in b) 
      is localized along the triangular orbit 
      $\overline{AAB}$. Again there are some deviations
      of the pattern of the eigenfunction near the symmetry axis.
      Also,  the momentum distributions do not show a clear
      enhancement in the momentum directions of the orbit.}
     {fig:kardi-two-D}

\afterpage{\clearpage}

The first eigenfunction displayed in fig.~\ref{fig:kardi-two-C}
shows localization along the cusp orbit $CAAC$,
and thus is an example of an {\it diffractive  scar}.
The enhanced momentum directions seen in $I_{1908}(\varphi)$ are
not precisely at the places of the directions of the
orbit $CAAC$, but of the same quality of the agreement
of the structure visible for $|\psi_{1908}(\BFx)|^2$
with the shape of the orbit.
For the second example shown in fig.~\ref{fig:kardi-two-C}
one observes that $|\psi_{1538}(\BFx)|^2$ is clearly
localized along the billiard boundary.
Due to the non-convexity of the billiard there are no whispering
gallery orbits as in the stadium billard. However,
as found in \cite{BaeDul97} in the cardioid there exist families 
of periodic orbits which accumulate in length. A candidate
responsible for the structure visible in $|\psi_{1538}(\BFx)|^2$
is the family of orbits with code $\overline{A^nBABB}$.
This type of orbit also accounts for the additional structures
visible in the interior of the billiard. As for an orbit running
along the boundary all momentum directions occur,
one expects to see only the two further directions
of the part of the orbit inside the billiard. Indeed,
there is an enhanced probability for these two directions,
which are marked by triangles in the plot of $I_{1538}(\varphi)$.
One should remark, that there is some additional structure
visible in $|\psi_{1538}(\BFx)|^2$ which could correspond to the
$\overline{AAABBB}$ orbit (see fig.~\ref{fig:kardi-two-A}).

We conclude our survey of eigenfunctions in momentum space
with two further examples of localized eigenfunctions,
see fig.~\ref{fig:kardi-two-D}. The first eigenfunction
is localized along the $\overline{AABB}$ orbit, which is also seen 
in the corresponding momentum distributions. Near
to the symmetry axis $|\psi_{1252}(\BFx)|^2$ the enhanced
region is much larger, presumably due to interference effects.
The eigenfunction with $n=1915$ of odd symmetry is localized
along the triangular orbit $\overline{AAB}$.
Near the symmetry line the  structures of the eigenfunction
deviate from the shape of the orbit, which could be caused 
by some contribution of the $CAAC$ orbit,
or again by interference effects.
Both directions of the $\overline{AAB}$ orbit are also seen
well in the pictures of the momentum distribution.

For the cardioid billiard
one generally observes a finer scale on wich $I_n$ fluctuates, 
compared to the stadium billiard. This is an effect of the stadium billiard, 
the long parallel segments of the boundary are reflected in the 
momentum distributions as a stretching of the structures 
in the $p_y$ direction. 
Therefore the radial integration acts as a smoothing except near 
$\varphi=0$ where accordingly $I_n(\varphi)$ fluctuates on a much 
finer scale.

\section{Summary, Applications and Discussion}

In this paper we have proposed the representation of
eigenfunctions in momentum space as an important
and useful complementary picture to the commonly used
position representation.
In particular we have introduced the radially integrated
momentum  distribution $I_n(\varphi)$, which captures 
all essential features of the momentum distribution
$|\widehat{\psi}_{n}(\BFp)|^2$.
As $I_n(\varphi)$ is just a one-dimensional function, 
it provides the information on the 
momentum distribution in a very condensed form.
Explicit formulae for $\widehat{\psi}_{n}(\BFp)$
and $I_n(\varphi)$, and also the angular integrated
momentum distribution $R_n(r)$, are given
in terms of the normal derivative $u_n(s)$
along the billiard boundary, which
allows for an efficient numerical computation of these quantities.

For the stadium and the cardioid billiard several
examples of  $|{\psi}_{n}(\BFq)|^2$,
$|\widehat{\psi}_{n}(\BFp)|^2$, $I_n(\varphi)$,
and  $R_n(r)$ are given. 
For the stadium billiard in particular the bouncing ball modes
lead to significant peaks in the radially integrated
momentum densities $I_n(\varphi)$. Also eigenfunctions
showing scarring in position space lead to scarred states in momentum
space. In addition to eigenstates showing localization
around unstable periodic orbits, we have found
for the cardioid billiard examples of eigenfunctions
which are localized along cusp orbits, which run into
the singularity of the cardioid billiard.
Such {\it diffractive scars} are surprisingly clearly visible,
despite the fact that the contribution of the corresponding
orbit to the density of states is of lower order in $\hbar$
than the periodic orbits \cite{BruWhe96}.
Our numerical studies also reveal that there are
eigenstates, for which the correspondence between localization
in position space and localization in momentum space
is not as clear as one would expect.
We see two main reasons for this: firstly, the fluctuations
appear to be larger in momentum space than in position space
(this is also seen in the results for the rate of quantum
ergodicity in momentum space,
which appears to be slower than in position space \cite{BaeSchSti98}).
Secondly, scars which are weak both in position and momentum space are 
better visible in position space. This is because the usually 
symmetric pattern 
of the scar can be  detected by the eye 
more easily against the fluctuating background 
in the two-dimensional plot of $|\psi_n(\BFx)|^2$, than in  
$|\widehat{\psi}_n(\BFp)|^2$
or  $I_n(\varphi)$. 

An application of 
the radially integrated momentum distribution 
$I_n(\varphi)$ could be 
to use it to construct a quantity
which detects localized eigenfunctions. A suitable
definition might be 
  $I_n^{\text{scar}} := \sum_{\varphi_i} \frac{1}{2\delta l_i} \int_{\varphi_i-\delta}^{\varphi_i+\delta} I_n(\varphi) \; \ud \varphi$,
where $l_i$ is the geometric length of the orbit segment 
corresponding to the direction $\varphi_i$ and $\delta$ is some
appropriate width, which might be chosen to be energy dependent.
In this context it may be also useful to employ the results
of the semiclassical studies of eigenfunctions \cite{Bog88,Ber89,AgaFis93}
to obtain expressions for $I_n(\varphi)$ in terms of periodic orbits.
A number of different scar measures have been defined and studied,
like for example the integral of the Wigner function over a tube 
around the orbit in phase space \cite{AgaFis93},
the integral over a tube in position
space \cite{LiHu98}, or the use of quantum
Poincar\'e sections \cite{SimVerSar97}, see \cite{Kap99} for a recent review 
and references therein.

The use of momentum distributions as representations
of eigenstates may also be useful for other systems,
like systems with potential etc.
The definition of the radially integrated momentum density
may easily be generalized to other scaling systems.
The usefulness of $I_n(\varphi)$ is due
to the fact that the classical motion occurs
on straight lines in Euclidean billiards. If this
does not hold, the above simple interpretation of $I_n(\varphi)$ is lost.

We believe that the use of momentum distributions, in
particular the radially integrated momentum distribution
is a convenient and useful representation,
providing additional information to the commonly 
used position space representation.

\vspace{1cm}

{\bf Acknowledgements}

We would like to thank 
Prof.\,F.\,Steiner 
for useful comments and 
Prof.\ M. Robnik and Dr.\ T. Prosen
for their kind provision of the eigenvalues of the cardioid billiard.  
The three--dimensional pictures in 
figs.~\ref{fig:stadium-ThreeD-A}, \ref{fig:stadium-ThreeD-B},
\ref{fig:cardid-ThreeD-A} and \ref{fig:cardid-ThreeD-B}
are visualized using {\tt Geomview} 
from {\it The Geometry Center} of the University of
Minnesota and then rendered using {\tt Blue Moon Rendering Tools} 
written by L.I.~Gritz.
A.B.\ acknowledges support by the 
Deutsche Forschungsgemeinschaft under contract No. DFG-Ste 241/7-3.


\vspace{1ex}

\section*{Appendix: Symmetry reduction}


When the billiard has some symmetries they can be used to 
reduce the domain of integration over the boundary to a smaller part 
in integrals as (\ref{Pn}). We will discuss as examples the 
stadium billiard and the cardioid billiard. To simplify 
the notation we use $\BFx =(x,y)$.

The stadium billiard has two symmetries, it is symmetric 
with respect to reflection at the $x$-axis and at the 
$y$-axis. Therefore the eigenfunctions can be choosen to be 
even or odd with respect to each of these symmetries. So we will 
treat the case of a function $\psi (x,y)$ which   satisfies 
\begin{align}
\psi (-x,y)&=(-1)^{\lx} \psi (x,y)\\ 
\psi (x,-y)&=(-1)^{\ly} \psi (x,y)\,\, ,
\end{align}
with $\lx ,\ly \in \{0,1\}$.  
Thus $\psi$ either satisfies Dirichlet or Neumann boundary conditions 
on the symmetry axes. The symmetry of the wavefunction implies now 
certain symmetries of the normal derivative 
$u(\omm )=\BFn(\omm)(\nabla \psi)(\BFx (\omm)) $. 
Denoting by $L$ the total length of the boundary of the full system, we can 
express $u(\omm )$ for $\omm \in [L/4,L/2]$, $\omm \in [L/2,3L/4]$ and 
$\omm \in [3L/4,L]$ in terms of  $u(\omm )$ for $\omm \in [0,L/4]$. Here 
we chose $s=0$ to be the intersection of the horizontal symmetry axis with 
the right semicircle, and $\omm$ is oriented   
counterclockwise. 
One gets for $\omm \in [0,L/4]$
\begin{align}
u(L/2-\omm )&=(-1)^{\lx}u(\omm ) \label{symm1} \\ 
u(L/2+\omm )&=(-1)^{\lx+\ly}u(\omm )  \label{symm2}  \\
u(L-\omm )&=(-1)^{\ly}u(\omm )\,\, . \label{symm3}
\end{align}
These relations can be used to reduce an integral over 
the boundary $\int_0^L f(\omm)u(\omm )\, \ud \omm $ to an integral 
over one quarter of the boundary. For instance with (\ref{symm1}) 
one gets 
\begin{equation}
\begin{split}
\Int_{L/4}^{L/2}u(\omm )f(\omm)\, \ud \omm 
&=\Int_{-L/4}^{0}u(L/2+\omm ) f(L/2+\omm)\, \ud \omm  \\
&=\Int_{0}^{L/4} u(L/2-\omm )f(L/2-\omm)\, \ud \omm 
=\Int_{0}^{L/4} (-1)^{\lx}u(\omm )f(L/2-\omm)\, \ud \omm \,\, ,
\end{split}
\end{equation}
and applying (\ref{symm2}) and (\ref{symm3})  to the other parts of the 
integral leads to 
\begin{equation}\label{symm_reduction}
\begin{split}
  \Int_0^L u(\omm )f(\omm)\, \ud \omm 
  =\Int_0^{L/4}u(\omm )[& f(\omm)+(-1)^{\lx}f(L/2-\omm) \\
  &+(-1)^{\lx+\ly}f(L/2+\omm) +(-1)^{\ly}f(L-\omm)]\, \ud \omm \,\, ,
\end{split}
\end{equation}
which is the desired reduction of the integral. 
We have to treat as well double integrals over the boundary of the 
form $\int_0^L \int_0^L u(\omm )\,\overline{u}(\omm ' )\, f(\omm ,\omm ')
 \; \ud\omm \,\ud\omm '$. If we use \eqref{symm_reduction} we get 
an integral with 16 terms, but for special $f(\omm ,\omm ')$ this 
can be reduced further. If $f$ is of the form 
\begin{equation}\label{dist_f}
f(\omm ,\omm ')=g(|\BFx(\omm )-\BFx(\omm ' )|)
\end{equation}
which is the case in our applications, then we have 
\begin{equation}\label{double_symm}
f(\omm ,\omm ')=f(L/2-\omm ,L/2-\omm ')=
f(L/2+\omm ,L/2+\omm ')=f(L-\omm ,L-\omm ')\,\, ,
\end{equation}
because of the symmetry of the billiard the distance between the 
points on the boundary is the same for all four pairs of arguments of $f$. 
Applying \eqref{double_symm} and  \eqref{symm_reduction} 
to a double integral over a function 
of the type \eqref{dist_f} leads to 
\begin{equation}
\Int_0^L \Int_0^L u(\omm )\,\overline{u}(\omm ' )\, f(\omm ,\omm ')
 \; \ud\omm \,\ud\omm ' =
\Int_0^{L/4} \Int_0^{L/4} u(\omm )\,\overline{u}(\omm ' )
\tilde{f}(\omm ,\omm ')\; \ud\omm \,\ud\omm '\,\, ,
\end{equation}
with
\begin{equation}
\begin{split}
\tilde{f}(\omm ,\omm ')=4
[ & f(\omm ,\omm ')+(-1)^{\lx}f(\omm ,L-\omm ') \\
  & +(-1)^{\lx+\ly}f(\omm ,L/2+\omm ')+(-1)^{\ly}f(\omm ,L/2-\omm ')]\,\, .
\end{split}
\end{equation}

The cardioid billiard has only one reflection symmetry at the 
$x$-axis, so the eigenfunctions can be chosen to be even or odd, 
\begin{equation}
\psi (x,-y)=(-1)^{\ly}\psi (x,y)
\end{equation}
with $\ly\in\{0,1\}$. For the normal derivative this leads to 
\begin{equation}
u(L-\omm )=(-1)^{\ly} u(\omm )\,\, ,
\end{equation}
and so we get for an integral over the boundary 
\begin{equation}
\Int_0^L u(\omm )f(\omm )\,\, \ud\omm =\Int_0^{L/2}u(\omm )\, 
[f(\omm )+(-1)^{\ly} f(L-\omm )]\,\, \ud\omm\,\, .
\end{equation}
For a double integral over a function of the type \eqref{dist_f} 
one obtains 
\begin{equation}
\Int_0^L \Int_0^L u(\omm )\,\overline{u}(\omm ' )\, f(\omm ,\omm ')
 \; \ud\omm \,\ud\omm ' =
\Int_0^{L/2} \Int_0^{L/2} u(\omm )\,\overline{u}(\omm ' )
2[f(\omm ,\omm ')+(-1)^{\ly} f(\omm ,L-\omm ')]\; \ud\omm \,\ud\omm '\,\, .
\end{equation}

\end{document}